\documentclass[amsfonts,letterpaper,aps,prb,twocolumn,groupedaddress,citeautoscript,floatfix]{revtex4}
\usepackage{graphicx} 
\usepackage{amsmath}
\usepackage{bm} 
\usepackage{dcolumn} 

\begin{document}

\title{Calculations of two-color interband optical injection and control of carrier population, spin, current, and spin current
in bulk semiconductors.}

\author{R. D. R. Bhat}
\author{J. E. Sipe}
\affiliation{Department of Physics and Institute for Optical
Sciences, University of Toronto, 60 St. George Street, Toronto,
Ontario M5S 1A7, Canada}

\date{December 28, 2005}

\begin{abstract}
Quantum interference between one- and two-photon absorption pathways
allows coherent control of interband transitions in unbiased bulk
semiconductors; carrier population, carrier spin polarization,
photocurrent injection, and spin current injection can all be
controlled. We calculate injection spectra for these effects using a
$14 \times 14$ $\mathbf{k}\cdot \mathbf{p}$ Hamiltonian including
remote band effects for five bulk semiconductors of zinc-blende
symmetry: InSb, GaSb, InP, GaAs, and ZnSe. Microscopic expressions
for spin-current injection and spin control accounting for spin
split bands are presented. We also present analytical expressions
for the injection spectra derived in the parabolic-band
approximation and compare these with the calculation nonperturbative
in $\mathbf{k}$.
\end{abstract}


\maketitle

\section{Introduction}

When a bulk semiconductor is simultaneously irradiated by an optical
field and its phase-coherent second harmonic, quantum interference
between one- and two-photon absorption pathways enables excitation
of carrier distributions with interesting properties
\cite{Atanasov96, Fraser99PRL, BhatSipe00, Stevens_pssb}. Such
excitation, even without an external bias, can produce ballistic
photocurrents \cite{Hache97}, spin-polarized currents
\cite{StevensJAP}, and pure spin currents
\cite{StevensPRL03,HubnerPRL03}. Characteristically of quantum
interference, these currents are sensitive to the phases of the two
optical fields. In noncentrosymmetric semiconductors, the phases can
also be used to control the total population of photoexcited
carriers \cite{Fraser99PRL}, and the net carrier spin polarization
\cite{Stevens_pssb, Stevens05_110}. Which of these effects occur
depends on the polarization states of the fields.

These are examples of ``$n+m$'' coherent control schemes, in which a
two-color light field controls a physical or chemical process by
interference of $n$- and $m$-photon transitions \cite{Manykin67,
ShapiroBrumer97, Gordon99}. In semiconductors, ``1+2'' excitation
has been discussed for impurity-band absorption \cite{Entin89}, free
carrier absorption \cite{Baskin88, Entin89, Alekseev99}, quantum
wells \cite{Dupont95, Potz98, Khurgin98, Najmaie03, Marti04,
MartiReview04, RumyantsevIQEC_04, Najmaie05a, Duc05}, and quantum
wires \cite{Marti05}, but our interest here is ``1+2'' coherent
control of interband transitions in unbiased bulk semiconductors
\cite{Atanasov96, Fraser99PRL, BhatSipe00, Stevens_pssb,
vanDrielSipeReview01, StevensReview04}. Such experiments have been
performed with either (a) two fields, typically short pulses, one
the generated second harmonic of the other \cite{Cote99, Cote03,
Fraser03, Fraser99PRL, HubnerPRL03, Hache97, HacheIEEE98,
Kerachian04, Roos03, StevensReview04, Stevens_pssb, StevensJAP,
Stevens05_110, StevensSST04, StevensJAP03, StevensPRL03}, or (b) a
single ultrashort pulse having at least an octave bandwidth
\cite{Fortier04, Roos05}.

Previous microscopic calculations of ``1+2'' processes in bulk
semiconductors fall into two categories: \textit{ab initio} density
functional methods have been used for current injection
\cite{Atanasov96} and population control \cite{Fraser99PRL}, while
simple analytical band models perturbative in $\mathbf{k}$ (with at
most eight spherical, parabolic bands) have been used for current
injection \cite{Atanasov96, Sheik-Bahae99, BhatSipe00, Kral00,
BhatSipeExcitonic05} and spin-current injection \cite{BhatSipe00}.
The former are best suited for excess energies on the order of eVs,
while the latter are only valid for excitation close to the band
edge and cannot be applied to population and spin control, which
vanish in such centrosymmetric models.

In this article, we calculate ``1+2'' processes using an
intermediate model that diagonalizes the $\mathbf{k}\cdot
\mathbf{p}$ Hamiltonian in a basis of 14 $\Gamma$-point states with
remote band effects included perturbatively. The model contains
empirically determined parameters \cite{PZ96, WinklerBook}.
Fourteen-band models (also called five-level models) have been used
to calculate band structures \cite{Rossler84, PZ90, MayerRossler91,
MayerRossler93b, PZ96}, linear \cite{MayerRossler93, BhatPRL05} and
non-linear \cite{HW94, Hutchings95, Hutchings97, Bhat_TPS_05}
optical properties, and spin decoherence properties \cite{Lau01,
Lau_condmat} of GaAs and other semiconductors. Winkler has recently
reviewed 14-band models \cite{WinklerBook}. The model is
nonperturbative in $\mathbf{k}$ and includes nonparabolicity,
warping, spin-splitting, and interband spin-orbit coupling. We apply
the 14-band model to the zinc-blende semiconductors InSb, GaSb, InP,
GaAs, and ZnSe.

We compare these results with analytic expressions derived in the
parabolic-band approximation (PBA) based on an expansion in
$\mathbf{k}$ about the $\Gamma$ point of $\mathbf{v}_{n,m}
(\mathbf{k})$, which is the matrix element governing optical
transitions. A one-photon transition is called ``allowed'' if the
zeroth-order term in its expansion is nonzero, and called
``forbidden'' otherwise. Two-photon transitions have two velocity
matrix elements, and thus have a hyphenated label depending on the
lowest-order terms in the expansions for each matrix element. For
example, if both matrix elements are independent of $\mathbf{k}$ to
lowest order, the two-photon transition is called
``allowed-allowed''. For current injection and spin-current
injection, we use expressions derived previously with an eight band
model \cite{BhatSipe00, BhatSipeExcitonic05}. For population control
and spin control, we derive expressions with the 14-band model.

The comparison between the PBA expressions and the numerical
calculation establishes an important microscopic difference between
current and spin-current control on the one hand, and population and
spin control on the other hand. Close to the band-gap, the former
result from the interference of allowed one-photon transitions and
allowed-forbidden two-photon transitions, whereas the latter result
from the interference of allowed one-photon transitions and
allowed-allowed two-photon transitions. This difference was posited
previously based on heuristic arguments \cite{Fraser03,
StevensReview04}.

Most of the early theory on semiconductor ``1+2'' processes
processes conceptually separated the optical injection of densities
and currents from the relaxation and transport of these quantities.
We follow this approach, and in this article, focus on microscopic
calculations of the optical injection. We note that relaxation and
transport have been studied with an effective circuit model
\cite{HacheIEEE98, Roos05}, hydrodynamic equations \cite{Atanasov96,
Cote03}, Boltzmann transport in the relaxation time approximation
\cite{HubnerPRL03}, a non-equilibrium Green function formalism
\cite{Kral00}, and the semiconductor Bloch equations \cite{Marti04,
MartiReview04, RumyantsevIQEC_04, Duc05}.

We model the optical field as a superposition of monochromatic
fields of frequency $\omega $ and $2\omega $:
\begin{equation}
\mathbf{E}(t) = \mathbf{E}_{\omega} \exp (-i\omega
t)+\mathbf{E}_{2\omega}\exp (-i2\omega t)+c.c. \label{eq:Efield}
\end{equation}
and we sometimes write $\mathbf{E}_{\omega / 2 \omega} = E_{\omega /
2 \omega} \mathbf{e}_{\omega / 2 \omega}$ and $E_{\omega / 2 \omega}
= |E_{\omega / 2 \omega}| \exp (i \phi_{\omega / 2 \omega})$. We
describe the fourteen-band model in Section \ref{sec:Model}, and use
it to study ``1+2'' current injection in Section \ref{sec:Current},
``1+2'' spin-current injection in Section \ref{sec:SpinCurrent},
``1+2'' population control in Section \ref{sec:Population}, and
``1+2'' spin control in Section \ref{sec:Spin}. We calculate the
injection of each ``1+2'' process using microscopic expressions
derived using velocity gauge ($\mathbf{A}\cdot\mathbf{v}$) coupling
in the long wavelength approximation, treating the field
perturbatively in the Fermi's golden rule limit, and using the
independent-particle approximation \cite{Atanasov96, Fraser99PRL,
BhatSipe00, footnoteIPAExcitonic}. For spin-current injection and
spin control, we use microscopic expressions that include the
coherence between spin-split bands. In Appendix \ref{App:kdepSO}, we
justify the neglect of $\mathbf{k}$-dependent spin-orbit coupling.
The parabolic-band approximation results are derived and discussed
in Appendix \ref{App:PBA_a-a}, and compared with the numerical
calculations in Sections \ref{sec:Current}--\ref{sec:Spin}. We
summarize and conclude in Section \ref{sec:Summary}.

\section{Model\label{sec:Model}}

\begin{figure}
\includegraphics{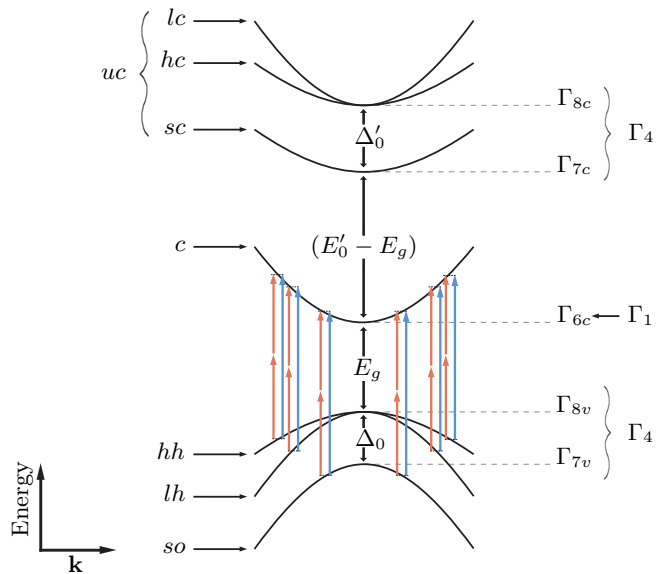}%
\caption{A schematic diagram of the fourteen-band model, indicating
band abbreviations (left), energies (center), symmetry of the
$\Gamma$-point states (right), and one- and two-photon transitions.
$\Gamma_{6}$, $\Gamma_{7}$, and $\Gamma_{8}$ indicate irreducible
representations of the $T_{d}$ double group, whereas $\Gamma_{1}$
and $\Gamma_{4}$ indicate irreducible representations of the $T_{d}$
point group. Note that spin-splitting, nonparabolicity, and warping
are not shown in this diagram.}\label{fig:14BandDiagram}
\end{figure}

The fourteen-band model Hamiltonian, which includes important
remote-band effects to order $k^{2}$, and which we denote $H_{14}$,
is given explicitly by Pfeffer and Zawadski \cite{PZ96,
footnotePZtypo}. The fourteen bands (counting one for each spin),
are shown in Fig.\ \ref{fig:14BandDiagram}. They comprise six
valence bands (two each for split-off, heavy and light holes) and
eight conduction bands (the two $s$-like ones at the band edge, and
the six next lowest ones which are $p$-like). We now briefly review
the derivation of $H_{14}$.

The one-electron field-free Hamiltonian is $H=H_{0}+H_{SO}$, where
$H_{0}=p^{2}/\left( 2m\right) +V$, the potential $V\left(
\mathbf{r}\right) $ has the symmetry of the crystal, and the
spin-orbit interaction $H_{SO}$ is
\[
H_{SO}=\frac{\hbar }{4m^{2}c^{2}}\bm{\sigma }\cdot \left( \bm{\nabla
} V\times \mathbf{p}\right) ,
\]
where $\bm{\sigma }$ is the dimensionless spin operator, $\bm{\sigma
}=2 \mathbf{S}/\hbar $. Note that relativistic corrections
proportional to $\left| \bm{\sigma }\times \bm{\nabla }V\right|
^{2}$ have been neglected \cite{LaxBook}. The eigenstates of $H$ are
Bloch states $\left| n\mathbf{k}\right\rangle $ with energy $\hbar
\omega _{n}\left( \mathbf{k}\right) $. The associated spinor wave
function $\phi _{n\mathbf{k}}\left( \mathbf{r}\right) \equiv \langle
\mathbf{r} | n\mathbf{k} \rangle $ can be written $\phi
_{n\mathbf{k}}\left( \mathbf{r}\right) =u_{n\mathbf{k}}\left(
\mathbf{r}\right) \exp \left( i\mathbf{k}\cdot \mathbf{r}\right) $,
where the spinor functions $u_{n\mathbf{k}}\left( \mathbf{r}\right)
$ have the periodicity of the crystal lattice. We use the notation
$\left| \overline{n \mathbf{k}}\right\rangle $ to denote the kets
for the $u$-functions; i.e.\ $ u_{n\mathbf{k}}\left(
\mathbf{r}\right) = \langle \mathbf{r} | \overline{n\mathbf{k}}
\rangle $. Note that $\left| \overline{n \mathbf{k}}\right\rangle
=\exp {\left( -i\mathbf{k}\cdot \mathbf{r}\right) } \left|
n\mathbf{k}\right\rangle $. The Hamiltonian for the $u$-function
kets, known as the $\mathbf{k}\cdot \mathbf{p}$ Hamiltonian, is
\cite{LaxBook, YuCardonaChapter2}
\[
H_{\mathbf{k}}=e^{-i\mathbf{k}\cdot \mathbf{r}}He^{i\mathbf{k}\cdot
\mathbf{r}}=H+\frac{\hbar ^{2}k^{2}}{2m}+\hbar \mathbf{k}\cdot
\mathbf{v},
\]
where the velocity operator $\mathbf{v}\equiv \left( i/\hbar \right)
\left[ H,\mathbf{r}\right] $ is
\begin{equation}
\mathbf{v}=\frac{1}{m}\mathbf{p}+\frac{\hbar }{4m^{2}c^{2}}\left(
\bm{\sigma }\times \bm{\nabla }V\right) .  \label{eq:v_operator}
\end{equation}
The second term in $\mathbf{v}$, the anomalous velocity, which leads
to $\mathbf{k}$-dependent spin-orbit coupling in $H_{\mathbf{k}}$,
can be neglected for the processes we consider as shown in Appendix
\ref{App:kdepSO}; in the rest of this article, we assume that it
vanishes.

The states $\left| n,\mathbf{k=\mathbf{0}}\right\rangle $ are a
complete set of eigenstates for the Hamiltonian $H$ on the space of
cell-periodic functions. Thus cell-periodic eigenstates of
$H_{\mathbf{k}}$ can be expanded in the infinite set of states
$\left| n,\mathbf{k=\mathbf{0}}\right\rangle $. The ``bare''
fourteen-band model truncates this expansion to a set of fourteen
states, corresponding to the fourteen bands closest in energy to the
fundamental band gap at the $\Gamma$ point \cite{PZ90}.

In a semiconductor of zinc-blende symmetry, the states $\left\{
\left| n, \mathbf{k=\mathbf{0}} \right\rangle  |n=1..14\right\}$ are
conveniently expanded in the eigenstates of $H_{0}$, $\left\{ \left|
S\right\rangle ,\left| X\right\rangle ,\left| Y\right\rangle ,\left|
Z\right\rangle ,\left| x\right\rangle ,\left| y\right\rangle ,\left|
z\right\rangle \right\} \otimes \left\{ \left| \uparrow
\right\rangle ,\left| \downarrow \right\rangle \right\} $, where,
under the point group $T_{d}$, $\left| S\right\rangle $ transforms
like $\Gamma _{1}$, $\left\{ \left| X\right\rangle ,\left|
Y\right\rangle ,\left| Z\right\rangle \right\} $ and $\left\{ \left|
x\right\rangle ,\left| y\right\rangle ,\left| z\right\rangle
\right\} $ transform like $\Gamma _{4}$ \cite{YuCardonaChapter2}.
The $\left\{ \left| \uparrow \right\rangle ,\left| \downarrow
\right\rangle \right\}$ comprises the usual spin $1/2$ states:
\begin{subequations}
\label{eq:Pauli}
\begin{align}
\left\langle \uparrow \right| \bm{\sigma }\left| \uparrow
\right\rangle &=-\left\langle \downarrow \right| \bm{\sigma }\left|
\downarrow
\right\rangle =\mathbf{\hat{z}} \\
\left\langle \uparrow \right| \bm{\sigma }\left| \downarrow
\right\rangle &= \left( \left\langle \downarrow \right| \bm{\sigma
}\left| \uparrow \right\rangle \right)^{*}=
\mathbf{\hat{x}}-i\mathbf{\hat{y}}.
\end{align}
\end{subequations}
The non-zero matrix elements of $\left( \bm{\nabla }V\times
\mathbf{p}\right) $ are
\begin{align*}
\left\langle X\right| \left( \bm{\nabla }V\times \mathbf{p}\right)
^{y}\left| Z\right\rangle &\equiv i\frac{4m^{2}c^{2}}{3\hbar }\Delta _{0}, \\
\left\langle x\right| \left( \bm{\nabla }V\times \mathbf{p}\right)
^{y}\left| z\right\rangle &\equiv i\frac{4m^{2}c^{2}}{3\hbar }\Delta _{0}^{\prime}, \\
\left\langle X\right| \left( \bm{\nabla }V\times \mathbf{p}\right)
^{y}\left| z\right\rangle &\equiv i\frac{4m^{2}c^{2}}{3\hbar
}\Delta^{-},
\end{align*}
cyclic permutations of these [e.g.\ $\left\langle x\right| \left(
\bm{\nabla }V\times \mathbf{p}\right) ^{y}\left| z\right\rangle
=\left\langle z\right| \left( \bm{\nabla }V\times \mathbf{p}\right)
^{x}\left| y\right\rangle =\left\langle y\right| \left( \bm{\nabla
}V\times \mathbf{p}\right) ^{z}\left| x\right\rangle $], and those
generated by Hermitian conjugation of these. The above equations
define the spin-orbit energies $\Delta _{0}$ and $\Delta
_{0}^{\prime}$, and the interband spin-orbit coupling $\Delta^{-}$
\cite{CPB65,CCF88}. The fourteen basis states $\left\{ \left|
n,\mathbf{k=\mathbf{0}} \right\rangle |n=1..14\right\}$ for $H_{14}$
are
\begin{subequations}
\label{eq:basis}
\begin{align}
\left| \Gamma _{7v},\pm 1/2\right\rangle &= \pm
\frac{1}{\sqrt{3}}\left| Z\right\rangle \left| \alpha _{\pm
}\right\rangle +\frac{1}{\sqrt{3}}\left|
X\pm iY\right\rangle \left| \alpha _{\mp }\right\rangle \\
\left| \Gamma _{8v},\pm 1/2\right\rangle &= \mp
\sqrt{\frac{2}{3}}\left| Z\right\rangle \left| \alpha _{\pm
}\right\rangle +\frac{1}{\sqrt{6}}\left|
X\pm iY\right\rangle \left| \alpha _{\mp }\right\rangle \\
\left| \Gamma _{8v},\pm 3/2\right\rangle &= \pm
\frac{1}{\sqrt{2}}\left|
X\pm iY\right\rangle \left| \alpha _{\pm }\right\rangle \\
\left| \Gamma _{6c},\pm 1/2\right\rangle &= i\left| S\right\rangle
\left|
\alpha _{\pm }\right\rangle \\
\left| \Gamma _{7c},\pm 1/2\right\rangle &= \pm
\frac{1}{\sqrt{3}}\left| z\right\rangle \left| \alpha _{\pm
}\right\rangle +\frac{1}{\sqrt{3}}\left| x\pm iy \right\rangle
\left|\alpha _{\mp}\right\rangle \\
\left| \Gamma _{8c},\pm 1/2\right\rangle &= \mp
\sqrt{\frac{2}{3}}\left| z\right\rangle \left| \alpha _{\pm
}\right\rangle +\frac{1}{\sqrt{6}}\left| x\pm iy \right\rangle
\left|\alpha _{\mp}\right\rangle \\
\left| \Gamma _{8c},\pm 3/2\right\rangle &= \pm
\frac{1}{\sqrt{2}}\left| x\pm iy\right\rangle \left| \alpha _{\pm
}\right\rangle ,
\end{align}
\end{subequations}
where $\left| \alpha _{+}\right\rangle =\left| \uparrow
\right\rangle $ and $\left| \alpha _{-}\right\rangle =\left|
\downarrow \right\rangle $. The states are labeled with their
transformation property under the double group for $T_{d}$, and with
a pseudo-angular momentum notation. In the basis \eqref{eq:basis},
$H_{\mathbf{k}=\mathbf{0}}$ is diagonal except for terms
proportional to $\Delta ^{-}$. The connection between the
eigenvalues of $H_{\mathbf{k}=\mathbf{0}}$ for the $\Gamma $-point
eigenstates and the eigenvalues of $H_{0}$ is given by Pfeffer and
Zawadski \cite{PZ90}. The nonzero matrix elements of momentum, which
appear in $H_{\mathbf{k}}$, are
\begin{subequations}
\label{eq:P_matrix}
\begin{align}
\left\langle S\right| p^{x}\left| X\right\rangle &= \left\langle
S\right| p^{y}\left| Y\right\rangle =\left\langle S\right|
p^{z}\left| Z\right\rangle
\equiv i m P_{0} / \hbar \\
\left\langle S\right| p^{x}\left| x\right\rangle &= \left\langle
S\right| p^{y}\left| y\right\rangle =\left\langle S\right|
p^{z}\left| z\right\rangle
\equiv i m P_{0}^{\prime} / \hbar \\
\begin{split}
\left\langle X\right| p^{y}\left| z\right\rangle &= \left\langle
Y\right| p^{z}\left| x\right\rangle =\left\langle Z\right|
p^{x}\left| y\right\rangle = \left\langle Z\right| p^{y}\left|
x\right\rangle \\
&= \left\langle Y\right| p^{x}\left| z\right\rangle =\left\langle
X\right| p^{z}\left| y\right\rangle  \equiv i m Q / \hbar .
\end{split}
\end{align}
\end{subequations}
Eq.\ \eqref{eq:P_matrix} defines the parameters $P_{0}$,
$P_{0}^{\prime}$, and $Q$. They are sometimes expressed as energies
$E_{P}$, $E_{P^{\prime}}$, and $E_{Q}$ with the connections $E_{P}=
2 m P_{0}^{2}/ \hbar ^{2}$, etc.

The ``bare'' fourteen-band model has eight empirical parameters
$E_{g}$, $\Delta_{0}$, $E_{0}^{\prime}$, $\Delta_{0}^{\prime}$,
$\Delta^{-}$, $P_{0}$, $Q$, and $P_{0}^{\prime}$. Its quantitative
accuracy is improved by adding important remote band effects to
order $k^{2}$ using L\"{o}wdin perturbation theory \cite{Lowdin51},
which adds $\mathbf{k}$-dependent terms to the truncated
$14\times14$ Hamiltonian so that its solutions better approximate
those of the full Hamiltonian \cite{PZ96}. The remote band effects
are governed by the parameters $\gamma _{1}$, $\gamma _{2}$, $\gamma
_{3}$, $F$, and $C_{k}$. The parameters $\gamma _{1} $, $\gamma
_{2}$, and $\gamma _{3}$ are modified Luttinger parameters that
account for remote band effects on the valence bands. They are
related to the usual Luttinger parameters $\gamma _{1L}$, $\gamma
_{2L}$, and $\gamma _{3L}$ by the couplings with $\Gamma _{6c}$,
$\Gamma _{7c}$, and $\Gamma _{8c}$ bands, which are already
accounted for in the ``bare'' fourteen-band model \cite{PZ96}:
\begin{align*}
\gamma_{1} &= \gamma_{1L} - \frac{E_{P}}{3 E_{g}} - \frac{E_{Q}}{3
E_{0}^{\prime}} - \frac{E_{Q}}{3
E_{0}^{\prime}+\Delta_{0}^{\prime}},
\\
\gamma_{2} &= \gamma_{2L} - \frac{E_{P}}{6 E_{g}} + \frac{E_{Q}}{6
E_{0}^{\prime}}, \\
\gamma_{3} &= \gamma_{3L} - \frac{E_{P}}{6 E_{g}} - \frac{E_{Q}}{6
E_{0}^{\prime}}.
\end{align*}
The parameter $F$ accounts for remote band effects on the lowest
conduction band, essentially fixing its effective mass to the
experimentally observed value. Finally, the parameter $C_{k}$ is the
small $\mathbf{k}$-linear term in the valence bands \cite{CCF88}.
The remote band effects can be removed by setting $\gamma _{1}=-1$
and $\gamma _{2}=\gamma _{3}=F=C_{k}=0$. The model includes neither
remote band effects on the $uc$ bands, nor remote band effects on
the $\Gamma _{6c}$-$\Gamma _{8v}$ and $\Gamma _{6c}$-$\Gamma _{7v}$
momentum matrix elements, although such terms exist in principle
\cite{WinklerBook}.

In summary, $H_{14}$ is a fourteen-band approximation to
$H_{\mathbf{k}}$ that incorporates some remote band effects. It can
be found in Eq.\ (5) of Pfeffer and Zawadzki, although with a
slightly different notation \cite{PZ96}. With their notation on the
left, and ours on the right: $E_{0}= - E_{g}$,
$E_{1}=E_{0}^{\prime}-E_{g}$, $\Delta_{1}=\Delta_{0}^{\prime}$,
$\overline{\Delta}=\Delta^{-}$, $P_{1}=P_{0}^{\prime}$. Also, our
$\Delta_{0}$ differs from theirs by a minus sign. Other authors have
also used different notations \cite{WinklerBook}. The fourteen bands
are shown schematically in Fig.\ \ref{fig:14BandDiagram} along with
the symmetry notation of the $\Gamma$-point states, and the notation
used to label the bands.

\subsection{Material parameters}

Numerical values for the thirteen parameters of the model are listed
in Table \ref{Table:parameters} for InSb, GaSb, InP, GaAs, and ZnSe.
They are taken from the literature, where they were chosen to fit
low-temperature experimental data. Of the two parameter sets
discussed by Pfeffer and Zawadzki for GaAs, we use the one
corresponding to $\alpha =0.085$ that they find yields better
results \cite{PZ96}. For InP, GaSb, and InSb, we use parameters from
Cardona, Christensen and Fasal \cite{CCF88}. For cubic ZnSe, we use
the parameters given by Mayer and Rossler \cite{MayerRossler93b}, we
use a calculated value of $C_{k}$ \cite{CCF88}, and we use $\Delta
^{-}=-0.238$ eV to give a $k^{3}$ conduction band spin-splitting
that matches the \textit{ab initio} calculation of Cardona,
Christensen and Fasal \cite{CCF88}. Winkler used these same
parameters for ZnSe, but took $\Delta ^{-}=0$ \cite{WinklerBook}.
There is more uncertainty in the parameters for ZnSe than in those
for the other materials \cite{MayerRossler93b}, but we include it as
an example of a semiconductor with a larger band gap.

\begin{table}[tbp]
\caption{Model parameters. }
\label{Table:parameters}%
\begin{ruledtabular}
\begin{tabular}{c|ddddd}
    &\multicolumn{1}{c}{GaAs}  & \multicolumn{1}{c}{InP}
     & \multicolumn{1}{c}{GaSb} & \multicolumn{1}{c}{InSb} &  \multicolumn{1}{c}{ZnSe}\\
\hline
$E_{g}$ (eV) & 1.519 & 1.424 & 0.813 & 0.235 & 2.820\\
$\Delta_{0}$ (eV) & 0.341 & 0.108 & 0.75 & 0.803 & 0.403\\
$E_{0}^{\prime}$ (eV) & 4.488 & 4.6 & 3.3 & 3.39 & 7.330\\
$\Delta_{0}^{\prime}$ (eV) & 0.171 & 0.50 & 0.33 & 0.39 & 0.090\\
$\Delta^{-}$ (eV) & -0.061 & 0.22 & -0.28 & -0.244 & -0.238\\
$P_{0}$ (eV\AA) & 10.30 & 8.65 & 9.50 & 9.51 & 10.628\\
$Q$ (eV\AA) & 7.70 & 7.24 & 8.12 & 8.22 & 9.845\\
$P_{0}^{\prime}$ (eV\AA) & 3.00 & 4.30 & 3.33 & 3.17 & 9.165\\
$\gamma_{1L}$ & 7.797 & 5.05 & 13.2 & 40.1 & 4.30\\
$\gamma_{2L}$ & 2.458 & 1.6 & 4.4 & 18.1 & 1.14\\
$\gamma_{3L}$ & 3.299 & 1.73 & 5.7 & 19.2 & 1.84\\
$F$ & -1.055 & 0 & 0 & 0 & 0\\
$C_{k}$ (meV\AA) & -3.4 & -14 & 0.43 & -9.2 & -14
\end{tabular}
\end{ruledtabular}
\end{table}

The parabolic-band approximation calculations use parameters from
Table \ref{Table:parameters}, and average effective masses derived
from the parameters in Table \ref{Table:parameters}.

\subsection{Matrix elements}

The relations between matrix elements of the Bloch states and
matrrix elements of the $u$-function kets are
\begin{gather}
\mathbf{v}_{nm}\left(\mathbf{k} \right) \equiv \left\langle
n\mathbf{k}\right| \mathbf{v} \left| m\mathbf{k} \right\rangle
=\left\langle \overline{n\mathbf{k}}\right| \mathbf{v}\left|
\overline{m\mathbf{k}}\right\rangle +\frac{\hbar
\mathbf{k}}{m}\delta _{nm}, \label{eq:v Bloch u}
\\
\left\langle n\mathbf{k} \right| \mathbf{S} \left|
m\mathbf{k}\right\rangle =\left\langle \overline{n\mathbf{k}}\right|
\mathbf{S}\left| \overline{m\mathbf{k}}\right\rangle
\\
\left\langle n\mathbf{k} \right| v^{i}S^{j} \left| m\mathbf{k}
\right\rangle =\left\langle \overline{n\mathbf{k}}\right|
v^{i}S^{j}\left| \overline{m\mathbf{k}}\right\rangle +\frac{\hbar
k^{i}}{m}\left\langle \overline{n\mathbf{k}}\right| S^{j}\left|
\overline{m\mathbf{k}} \right\rangle .
\end{gather}

The matrix elements of the velocity operator, $\mathbf{v}$,
neglecting the anomalous velocity as discussed in Appendix
\ref{App:kdepSO}, can be calculated using \eqref{eq:v_operator},
\eqref{eq:P_matrix}, and the right side of \eqref{eq:v Bloch u}. The
matrix elements of the spin operator $\mathbf{S}$, can be found from
Eq.\ \eqref{eq:Pauli}. The matrix elements of $v^{i}S^{j}$ can be
similarly found in the basis of eigenstates of $H_{0}$. Each of
these can then be rotated to the basis \eqref{eq:basis} in which the
states $\left| \overline{m\mathbf{k}}\right\rangle$ are expanded.

It is well known that, in a crystal, $\mathbf{v}_{nn}\left(
\mathbf{k}\right) = \nabla _{\mathbf{k}}\omega _{n}\left(
\mathbf{k}\right) $. More generally,
\begin{equation} \label{eq:v nablaH identity}
\mathbf{v}_{nm}\left( \mathbf{k}\right) = \nabla_{\mathbf{k}}
\left\langle n \mathbf{k} \right| H \left| m \mathbf{k}\right\rangle
= \left\langle \overline{n \mathbf{k}} \right| \nabla_{\mathbf{k}}
H_{\mathbf{k}} \left| \overline{m \mathbf{k}}\right\rangle.
\end{equation}
These identities can be proven from the definitions
$H_{\mathbf{k}}=e^{-i\mathbf{k}\cdot \mathbf{r}}He^{i\mathbf{k}\cdot
\mathbf{r}}$ and $\mathbf{v}=\left( i/\hbar \right) \left[
H,\mathbf{r}\right]$, even for a non-local Hamiltonian. But when
remote band effects are included in a finite band model, they no
longer hold. That is, $\mathbf{v}_{nm}\left( \mathbf{k}\right)$
calculated using \eqref{eq:v Bloch u} and eigenstates of $H_{14}$ is
not equal to $\left\langle \overline{n \mathbf{k}} \right|
\nabla_{\mathbf{k}} H_{14} \left| \overline{m
\mathbf{k}}\right\rangle$. We explicitly restore these identities by
using $\left\langle \overline{n \mathbf{k}} \right|
\nabla_{\mathbf{k}} H_{14} \left| \overline{m
\mathbf{k}}\right\rangle$ to calculate $\mathbf{v}_{nm}\left(
\mathbf{k}\right)$. This approach can be described as including
remote band effects in the velocity operator. It was used for an
eight band calculation of linear absorption by Enders et al
\cite{Enders95}. This step is not critically important for the
effects calculated here, since remote band effects are generally
small.

\subsection{$\mathbf{k}$-space integration}

The optical calculations in this article have the form $\Theta $,
where
\begin{equation} \label{eq:GeneralOpticalCalc}
\Theta =\sum_{c,v}\int d^{3}kf_{cv}\left( H_{\mathbf{k}}\right)
\delta \left( \hbar \omega _{cv}\left( \mathbf{k}\right)
-2\hbar\omega\right) ,
\end{equation}
where $f_{cv}$ depends on matrix elements and energies of
eigenstates of $H_{\mathbf{k}}$, and where $\omega _{nm}\left(
\mathbf{k}\right) \equiv \omega _{n}\left( \mathbf{k}\right) -\omega
_{m}\left( \mathbf{k}\right) $. The integral in
\eqref{eq:GeneralOpticalCalc} is understood to be restricted to the
first Brillioun Zone, but we do not actively enforce the
restriction, since the photon energies considered here cause
transitions well within the first Brillioun Zone. Writing
$\mathbf{k}=\left( k_{cv},\theta _{\mathbf{k}},\phi
_{\mathbf{k}}\right) $ in spherical coordinates, where $k_{cv}$ is
the solution to
\begin{equation}
\hbar \omega _{cv}\left( k_{cv},\theta _{\mathbf{k}},\phi
_{\mathbf{k}}\right) -2\hbar \omega=0,  \label{eqn:RootFinding}
\end{equation}
we have
\begin{equation}
\Theta =8\sum_{c,v}\int_{0}^{\pi /2}\int_{0}^{\pi
/2}\frac{k_{cv}^{2}\sin \theta _{\mathbf{k}} f_{cv}\left(
H_{\mathbf{k}}\right) }{\left| \hbar \left( \mathbf{v}_{cc}\left(
\mathbf{k}\right) -\mathbf{v}_{vv}\left( \mathbf{k} \right) \right)
\cdot \mathbf{\hat{k}}\right| }d\phi _{\mathbf{k}}d\theta
_{\mathbf{k}} \label{eqn:BZIntegration},
\end{equation}
where we have used $\bm{\nabla }\omega _{n}\left( \mathbf{k}\right)
= \mathbf{v}_{nn}\left( \mathbf{k}\right) $ and the cubic symmetry
of the crystal. It is numerically convenient to do the sum over any
degenerate bands before the integral over $\theta _{\mathbf{k}}$ and
$\phi _{\mathbf{k}}$.

\subsection{Approximations\label{approximations_defined}}

The calculations of ``1+2'' effects in the following sections are
primarily labeled by the Hamiltonian used to approximate
$H_{\mathbf{k}}$. The complete fourteen-band model is denoted
$H_{14}$. The bare fourteen-band model, denoted
$H_{14\text{-Bare}}$, is $H_{14}$ without remote band effects. The
$8\times 8$ subset of the fourteen band Hamiltonian within the basis
$\left\{ \Gamma _{6c},\Gamma _{8v},\Gamma _{7v}\right\} $ is denoted
$H_{8}$. The spherical eight-band model, denoted $H_{8\text{Sph}}$,
is derived from $H_{8}$ by setting $ C_{k}=0$ and replacing $\gamma
_{2}$ and $\gamma _{3}$ by $\tilde{\gamma} \equiv \left( 2\gamma
_{2}+3\gamma _{3}\right) /5$;\cite{Baldereschi73} it is a spherical
approximation to the Kane model including remote band effects
\cite{Kane57}. The aforementioned calculations are non-perturbative
in $\mathbf{k}$; that is, in each case, the Hamiltonian is solved
numerically at each $\mathbf{k}$. The perturbative calculations of
Appendix \ref{App:PBA_a-a} are denoted PBA (parabolic-band
approximation).

The microscopic expression for each of the ``1+2'' effects contains
a sum over intermediate bands, which originates from the two-photon
amplitude. Unless otherwise noted, calculations include all possible
intermediate bands (eg., $H_{14}$ includes fourteen intermediate
bands, and $H_{8\text{Sph}}$ includes eight intermediate bands).
Calculations that restrict this sum are secondarily labeled to
reflect the restriction. The label ``$H_{14}$, no $uc$'' uses
$H_{14}$, but does not include $uc$ bands as intermediate states.
The label ``$H_{14}$, no $uc$/$so$'' uses $H_{14}$, but includes
neither $uc$ nor $so$ bands as intermediate states. The label
``$H_{14}$, 2BT'' uses $H_{14}$, but only includes two-band terms
(terms for which the intermediate band is the same as the initial or
final band). Similar labels are used for $H_{8\text{Sph}}$, for
example, ``$H_{8\text{Sph}}$-PBA, no so'' uses the perturbative
solution to $H_{8\text{Sph}}$ and does not include $so$ intermediate
states.

\section{Current\label{sec:Current}}

The current injection rate due to the field \eqref{eq:Efield} can be
written
\begin{equation}
\dot{J}^{i} = \eta _{\left( 1\right) }^{ijk}E_{2\omega}^{j*}
E_{2\omega}^{k} + \dot{J}_{(I)}^{i} + \eta _{\left( 2\right)
}^{ijklm}E_{\omega}^{j*}E_{\omega}^{k*}E_{\omega}^{l}E_{\omega}^{m},
\label{eq:phenom_current}
\end{equation}
where $\mathbf{J}$ is the macroscopic current density, and
\begin{equation}
\dot{J}_{(I)}^{i} = \eta_{(I)}
^{ijkl}E_{\omega}^{j*}E_{\omega}^{k*}E_{2\omega}^{l}+c.c.
\label{eq:phenom_current_I}
\end{equation}
The third rank tensor $\eta _{\left( 1\right) }^{ijk}$ describes
one-photon current injection (the circular photogalvanic effect
\cite{SturmanFridkin, GanichevReview}), the fifth rank tensor $\eta
_{\left( 2\right) }^{ijklm}$ describes two-photon current injection,
and the fourth rank tensor $\eta_{(I)} ^{ijkl}$ describes ``1+2''
current injection \cite{Atanasov96}. Aversa and Sipe showed that
$\eta_{(I)} ^{ijkl}$ is related to a doubly divergent part of the
third-order nonlinear susceptibility $\chi ^{\left( 3\right) }$
\cite{Aversa96}. In cubic materials with point group symmetry
$T_{d}$, $O_{h}$ or $O$, a general fourth rank tensor has four
independent components, but due to the intrinsic symmetry
$\eta_{(I)} ^{ikjl}=\eta_{(I)} ^{ijkl}$, $\eta_{(I)} $ has only
three independent components; there are 21 non-zero components of
$\eta_{(I)} $ in the standard cubic basis: $\eta
_{(I)}^{aaaa}=\eta_{(I)} ^{bbbb}=\eta_{(I)} ^{cccc}$, $\eta_{(I)}
^{baab}=\eta_{(I)} ^{abba}=\eta_{(I)} ^{caac}=\eta_{(I)}
^{acca}=\eta_{(I)} ^{cbbc}=\eta_{(I)} ^{bccb}$, and $\eta_{(I)}
^{a\left( ab\right) b}=\eta_{(I)} ^{b\left( bc\right) c}=\eta_{(I)}
^{c\left( ca\right) a}=\eta_{(I)} ^{a\left( ac\right) c}=\eta_{(I)}
^{c\left( cb\right) b}=\eta_{(I)} ^{b\left( ba\right) a}$ (the
components in parentheses can be exchanged), where $a$, $b$, and $c$
denote components along the principal cubic axes \cite{Atanasov96}.
This can be written
\begin{equation}
\eta _{(I)}^{ijkl}=i\frac{\eta _{B1}}{2}\left( \delta ^{ij}\delta
^{kl}+\delta ^{ik}\delta ^{jl}\right) +i\eta _{B2}\delta ^{il}\delta
^{jk} +i\eta _{C}\delta ^{ijkl}, \label{eq:phenom_eta_form}
\end{equation}
where $\delta ^{ij}$ is a Kronecker delta and the only non-isotropic
part is $\delta ^{ijkl}$, which we define in the principal cubic
basis as $\delta ^{ijkl}=1$ when $i=j=k=l$ and zero otherwise. The
three independent components are $\eta _{B1}\equiv -2i\eta ^{aabb}$,
$\eta _{B2}\equiv -i\eta ^{abba}$, and $\eta _{C}\equiv 2i\eta
^{aabb}+i\eta ^{abba}-i\eta ^{aaaa}$. Thus, in a cubic material,
\begin{equation}
\begin{split}
\dot{J}_{(I)}^{i} =& i\eta _{B1}\left( \mathbf{E}_{\omega}^{*}\cdot
\mathbf{E}_{2\omega}\right) E_{\omega}^{i*}+i\eta _{B2}\left(
\mathbf{E}_{\omega}\cdot \mathbf{E}_{\omega}\right)
^{*}E_{2\omega}^{i} \\
&+i\eta _{C}\delta
^{ijkl}E_{\omega}^{j*}E_{\omega}^{k*}E_{2\omega}^{l}+c.c.
\end{split}
\end{equation}
This generalizes the notation we used previously for a calculation
in the parabolic-band approximation \cite{BhatSipe00}, with the
connection $\eta _{B1}=eDB_{1}/\hbar $, and $\eta
_{B2}=eDB_{2}/\hbar $. In that, or any other spherical
approximation, $\eta _{C}=0$.

To calculate $\eta_{(I)} $, we use the microscopic expression first
given by Atanasov et al.\ \cite{Atanasov96}, modified to explicitly
include the sum over spin states
\cite{vanDrielSipeReview01,Najmaie03}. An alternate microscopic
expression has been derived in the length gauge \cite{Aversa96}, but
it has not yet been used in a calculation. In the independent
particle approximation that we employ here, $\eta_{(I)} $ is purely
imaginary \cite{Atanasov96} and hence $\eta _{B1}$, $\eta _{B2}$,
and $\eta _{C}$ are real, although they can be complex if excitonic
effects are included \cite{BhatSipeExcitonic05}.

\begin{figure}
\includegraphics[]{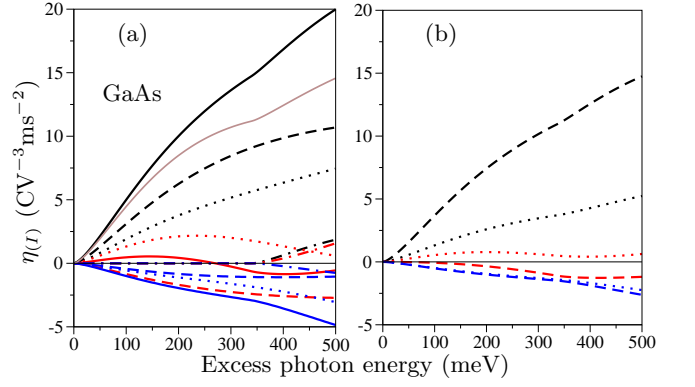}%
\caption{(color online): Spectra of $\eta _{B1}$ (black lines),
$\eta _{B2}$ (red lines), and $\eta _{C}$ (blue lines) for GaAs.
Panel (a) shows the contributions from each initial valence band;
dashed, dotted, and dashed-dotted lines include only transitions
from the $hh$, $lh$, and $so$ bands respectively, while the solid
lines include all three transitions. The thin solid, light brown
line in (a) is the total $\mathrm{Re} \left( \eta _{(I)}^{aaaa}
\right)$. Panel (b) separates the total into electron (dashed) and
hole (dotted) contributions.\label{fig:currentGaAs_initial}}
\end{figure}

The spectra of $\eta _{B1}$, $\eta _{B2}$, and $\eta _{C}$,
calculated for GaAs, are shown in Fig.\
\ref{fig:currentGaAs_initial}(a) along with the contributions to
each tensor component from each possible initial valence band. For a
given photon energy, electrons photoexcited from the $hh$ band have
higher energies and velocities than electrons photoexcited from the
$lh$ band; hence the dominant component $\eta _{B1}$ is larger for
$hh$-$c$ transitions than $lh$-$c$ transitions. The smallness of
$\eta _{B2}$ is due to contributions from the $hh$-$c$ transitions
having opposite sign to the $lh$-$c$ transitions, as shown
previously in the PBA \cite{BhatSipe00}.

Figure \ref{fig:currentGaAs_initial}(b) separates each tensor
component into an electron contribution and a hole contribution
(denoted $\eta _{e}$ and $\eta _{h}$ by Atanasov et al
\cite{Atanasov96}). Electrons make a larger contribution to $\eta
_{B1}$ than holes, due to the lower effective mass (and hence higher
velocity) of an electron than of a hole (much lower, in the case of
a heavy hole) with the same crystal momentum. Holes dominate $\eta
_{B2}$ at lower photon energies, while electrons dominate $\eta
_{B2}$ at higher energies. Both electrons and holes contribute
equally to the anisotropic component $\eta _{C}$.

\begin{figure*}
\includegraphics[]{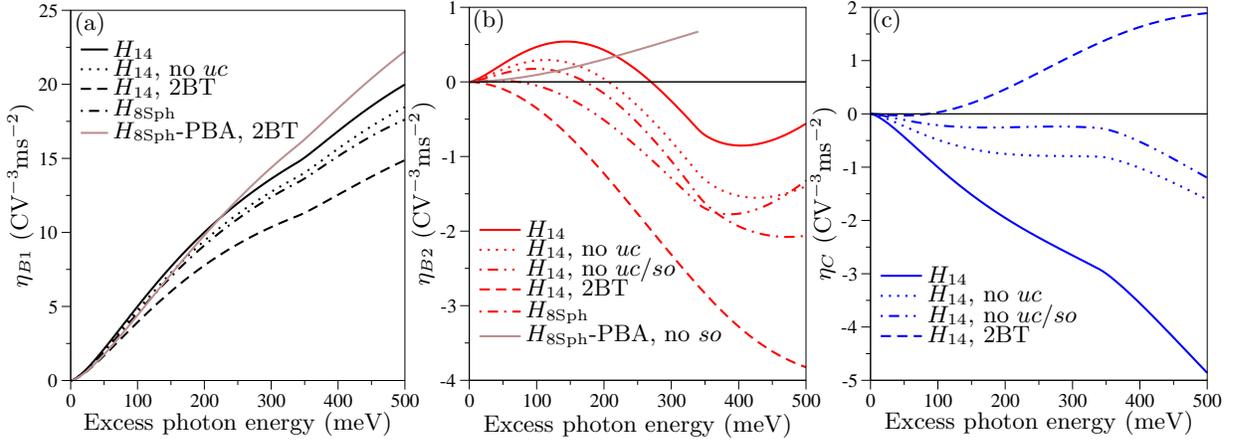}%
\caption{(color online): Approximations for GaAs current injection
tensor components (a) $\eta _{B1}$, (b) $\eta _{B2}$, and (c) $\eta
_{C}$. The approximations are described in Sec.\
\ref{approximations_defined}.\label{fig:currentGaAs_models}}
\end{figure*}

To help in understanding the importance of the various intermediate
states, in Fig.\ \ref{fig:currentGaAs_models} we compare the
calculated current injection tensor elements with various degrees of
approximation described in Sec.\ \ref{approximations_defined}.

The component $\eta _{B1}$ (and hence $\eta_{(I)} ^{aaaa}$, since
$\eta _{B1}$ is larger than $\eta _{B2}+\eta _{C}$) is dominated by
two-band terms. Three-band terms cause an increase, by as much as
34\%, of $\eta _{B1}$ [the difference between the dashed and solid
black lines in Fig.\ \ref{fig:currentGaAs_models}(a)]. Although not
shown in Fig.\ \ref{fig:currentGaAs_models}, most of the increase is
due to three-band terms with the $so$ band as an intermediate state.
Terms with the $uc$ bands as intermediate states only cause a small
increase to $\eta _{B1}$ (the difference between the dotted and
solid black lines). The warping of the bands is clearly not
important for $\eta _{B1}$, since the calculation with
$H_{8\text{Sph}}$ closely approximates the calculation ``$H_{14}$,
no $uc$'', which includes the same intermediate states.
Surprisingly, the ``$H_{8\text{Sph}}$-PBA, 2BT'' result
\cite{BhatSipe00, BhatSipeExcitonic05} closely approximates the
complete, non-perturbative fourteen-band calculation, even at excess
photon energies for which band nonparabolicity is significant. This
is due to a fortuitous compensation between the neglect of
nonparabolicity and the neglect of three-band terms. The
compensation is not as complete for all materials.

The component $\eta _{B2}$, which determines the current due to
orthogonal linearly polarized fields, is less forgiving to
approximations than the component $\eta _{B1}$. We have already seen
in Fig.\ \ref{fig:currentGaAs_initial} that $\eta _{B2}$ is small
due to a near cancellation of $hh$ and $lh$ initial states.
Reasonable accuracy on $\eta _{B2}$ thus requires higher accuracy on
the contribution from each initial state. In particular, three-band
terms must not be neglected. By comparing the dashed-dotted and
solid lines in Fig.\ \ref{fig:currentGaAs_models}(b), it can be seen
that, whereas the sum of the two-band terms is negative, the sum of
the three-band terms is positive and of the same magnitude. It is
useful to divide the three-band terms into three groups: those with
intermediate state from the $hh$ or $lh$ bands, those with
intermediate state from the $so$ band, and those with intermediate
state from one of the $uc$ bands. We find that each group
contributes roughly the same positive amount to $\eta _{B2}$ for
excess photon energies less than $\Delta _{0}$. The groups are added
successively to the 2BTs in the dashed, dotted, and solid lines in
Fig.\ \ref{fig:currentGaAs_models}(b). Three-band terms with $so$
intermediate states are less important at the higher excess photon
energies in Fig.\ \ref{fig:currentGaAs_models}(b). The warping of
the bands makes a small but non-negligible contribution to $\eta
_{B2}$, as seen in the difference between the dashed-double-dotted
and dotted lines of Fig.\ \ref{fig:currentGaAs_models}(b). The solid
brown line in Fig.\ \ref{fig:currentGaAs_models}(b) is the
``$H_{8\text{Sph}}$-PBA, no $so$'' result \cite{BhatSipe00}. At low
excess photon energies, it greatly underestimates $\eta _{B2}$ due
to the neglect of $so$ and $uc$ intermediate states, while at excess
photon energies greater than 100 meV, this is partly compensated for
by the neglect of nonparabolicity. It appears from the difference
between ``$H_{8\text{Sph}}$-PBA, no $so$'' and ``$H_{14}$, no
$uc$/$so$'' in Fig.\ \ref{fig:currentGaAs_models}(b) that
nonparabolicity becomes important at energies above 70 meV.

The term $\eta _{C}$ is purely due to cubic anisotropy by
definition; in any model that is spherically symmetric it is
identically zero. There is no cubic anisotropy in the ``bare''
(i.e.\ without remote band effects) eight-band model on the set
$\left\{ \Gamma _{6c},\Gamma _{8v},\Gamma _{7v}\right\} $. Cubic
anisotropy in the fourteen-band model is due to the momentum matrix
elements governed by the parameters $E_{Q}$ and $E_{P^{\prime }}$,
the interband spin-orbit coupling $\Delta ^{-}$, and remote bands
through $(\gamma _{2}-\gamma _{3})$ and $C_{k}$. From Fig.\
\ref{fig:currentGaAs_models}(c), it can be seen that three-band
terms are important for $\eta _{C}$. In fact, with only 2BTs
included, $\eta _{C}$ is positive for GaAs, whereas it is negative
with all terms included. From Fig.\ \ref{fig:currentGaAs_models}(c)
it can also be seen that the $so$ band and $uc$ bands are important
as intermediate states for $\eta _{C}$.

Our calculation of $\eta_{(I)} $ is of the same order of magnitude
as the \textit{ab initio} calculation of Atanasov et al.\
\cite{Atanasov96}, but its spectral dependence is different. In
particular, $\eta _{B1}$ agrees more closely with the PBA
calculation, as seen in Fig.\ \ref{fig:currentGaAs_models}(a).
Atanasov et al.\ had attributed the difference between their
\textit{ab initio} and PBA calculations to the assumption of
$\mathbf{k}$-independent velocity matrix elements in the PBA
\cite{Atanasov96}. However, our calculation accounts for the
$\mathbf{k}$-dependence of velocity matrix elements and agrees
closely (for $\eta _{B1}$ and $\mathrm{Re}\eta ^{aaaa}$) to the PBA.
The earlier \textit{ab initio} calculation \cite{Atanasov96} was, in
fact, inaccurate at low photon energies due to various computational
issues; an improved \textit{ab initio} calculation agrees with the
spectral dependence at low photon energy given here
\cite{NastosPrivateCommunication}.

\begin{figure*}
\includegraphics[]{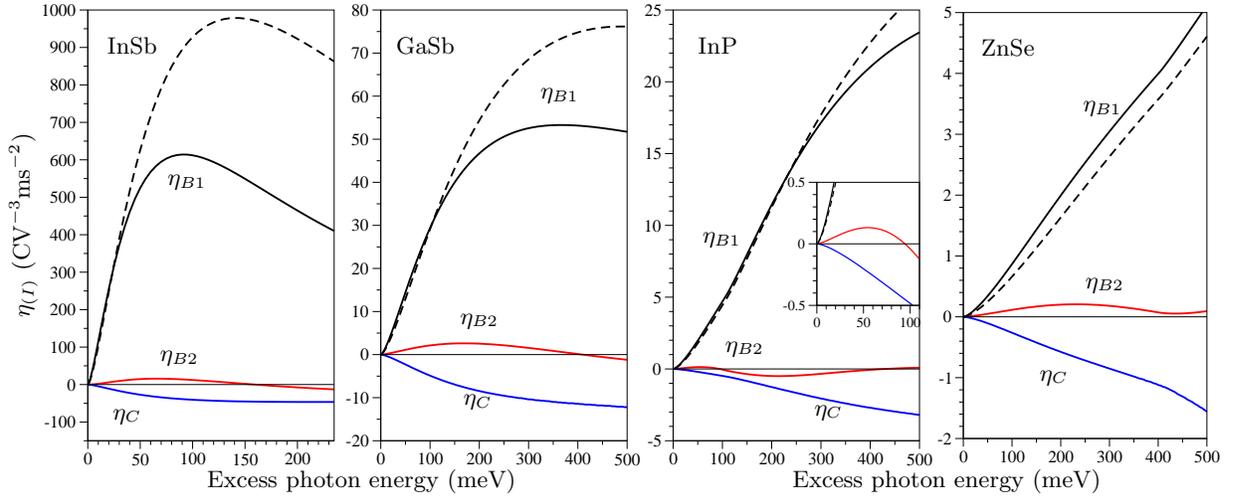}%
\caption{(color online): $\eta _{B1}$ (black), $\eta _{B2}$ (red),
and $\eta _{C}$ (blue) for (a) InSb, (b) GaSb, (c) InP, and (d)
ZnSe. The solid lines are calculated with the complete fourteen-band
model. The dashed line for $\eta _{B1}$ is ``$H_{8\text{Sph}}$-PBA,
2BT'' \cite{BhatSipe00,BhatSipeExcitonic05}. The inset of panel (c)
shows the area near the origin in more
detail.\label{fig:current_4materials}}
\end{figure*}

Figure \ref{fig:current_4materials} shows the spectra of $\eta
_{B1}$, $\eta _{B2}$, and $\eta _{C}$ calculated with $H_{14}$ for
InSb, GaSb, InP, and ZnSe. The dashed black line in Fig.\
\ref{fig:current_4materials} is the PBA result \cite{BhatSipe00,
BhatSipeExcitonic05}. The PBA appears to be a reasonable
approximation to $\eta _{B1}$ for excess energies less than about
$0.2E_{g}$. In each material, $\eta _{B2}\ll \eta _{B1}$, and in
each material except for ZnSe, the sign of $\eta _{B2}$ varies as a
function of frequency. The component $\eta _{C}$, which arises due
to cubic anisotropy, is negative for each material.

The cubic anisotropy of current injection due to colinearly
polarized fields can be significant enough that it should be
measurable. For fields colinearly polarized along
$\mathbf{\hat{e}}$, specified by polar angles $\theta $ and $\phi $
relative to the cubic axes,
\begin{equation}
\mathbf{\dot{J}}_{(I)} \cdot \mathbf{\hat{e}}=2\mathrm{Im}\left(
E_{\omega}^{2}E_{2\omega}^{*}\right) \left[ \eta _{B1}+\eta
_{B2}+\eta _{C}-\frac{\eta _{C}}{2}f\left( \theta ,\phi \right)
\right] ,
\end{equation}
where $f\left( \theta ,\phi \right) =\sin ^{2}\left( 2\theta \right)
+\sin ^{4}\left( \theta \right) \sin ^{2}\left( 2\phi \right) $. In
general, $\mathbf{\dot{J}}_{(I)}$ also has a component perpendicular
to $\mathbf{\hat{e}}$ that is proportional to $\eta _{C}$, but it
vanishes for $\mathbf{\hat{e}}$ parallel to $\left \langle
001\right\rangle $, $\left \langle 110\right\rangle $, $\left\langle
111\right\rangle $. The field polarization that maximizes the
current injection depends on the relative sign of $\eta _{C}$ and
$\mathrm{Re}\eta ^{aaaa}=\eta _{B1}+\eta _{B2}+\eta _{C}$. When they
have the opposite sign, current injection is a minimum for
$\mathbf{\hat{e}\parallel }\left \langle 001\right \rangle $ ($f=0$)
and a maximum for $\mathbf{\hat{e}\parallel }\left \langle 111\right
\rangle$ ($f=4/3$); for light normally incident on a $\left\{
001\right\} $ surface, the largest current injection occurs when
$\mathbf{\hat{e}\parallel }\left \langle 110\right \rangle$ ($f=1$).
When they have the same sign, current-injection is a maximum for
$\mathbf{\hat{e}\parallel }\left \langle 001\right \rangle $ and a
minimum for $\mathbf{\hat{e}\parallel }\left \langle 111\right
\rangle$. From the GaAs results shown in Fig.\
\ref{fig:currentGaAs_initial}(a), the current injection for the
three cases $\mathbf{\hat{e}\parallel }\left \langle 001\right
\rangle$, $\mathbf{\hat{e}\parallel }\left \langle 110\right
\rangle$, and $\mathbf{\hat{e}\parallel }\left \langle 111\right
\rangle$ are in the ratio 1 to 1.14 to 1.20 at the band edge, 1 to
1.15 to 1.20 at 200 meV excess photon energy, and 1 to 1.22 to 1.29
at 500 meV excess photon energy. In contrast, the \textit{ab initio}
calculation of Atanasov et al.\ yields larger ratios, for example 1
to 1.32 to 1.43 at 300 meV excess photon energy \cite{Atanasov96}.
This disagreement is consistent with the inaccuracy of the
\textit{ab initio} calculation discussed above. Initial experiments
with GaAs used $\mathbf{\hat{e}\parallel }\left[ 001\right] $
\cite{Hache97,StevensJAP}, whereas Roos et al.\ exploited the larger
signal for $\mathbf{\hat{e} \parallel }\left[ 110\right] $
\cite{Roos03}. For each of the materials shown in Fig.\
\ref{fig:current_4materials}, the minimum current injection is for
$\mathbf{\hat{e}\parallel } \left \langle 001\right \rangle$. It is
worth noting that two-photon absorption is also a minimum with
$\mathbf{\hat{e}\parallel }\left \langle 001\right \rangle$ for many
semiconductors \cite{Dvorak94, Hutchings94, Murayama95}. It seems
that both ``1+2'' current injection and two-photon absorption with
linearly polarized fields  are larger for $\mathbf{\hat{e}}$
directed along the bonds.

The cubic anisotropy of ``1+2'' current injection is pronounced for
cross-linearly polarized fields and opposite-circularly polarized
fields. For example, for cross-linearly polarized fields normally
incident on $\left( 001\right) $ with $\mathbf{\hat{e}}_{\omega} =
\mathbf{\hat{a}} \cos \phi + \mathbf{\hat{b}} \sin \phi $ and
$\mathbf{\hat{e}}_{2\omega} = - \mathbf{\hat{a} } \sin \phi +
\mathbf{\hat{b}} \cos \phi $,
\begin{equation}
\begin{split}
\mathbf{\dot{J}}_{(I)} =& \mathrm{Im}\left(
E_{\omega}^{2}E_{2\omega}^{*}\right) \\
&\times \left[ \left( 2\eta _{B2}+\eta _{C}\sin ^{2}\left( 2\phi
\right) \right) \mathbf{\hat{e}}_{2\omega}-\frac{\eta _{C}}{2}\sin
\left( 4\phi \right) \mathbf{\hat{e}}_{\omega}\right] .
\end{split}
\end{equation}
For fields with opposite circular polarizations, the current
injection is proportional to $\eta _{C}$ and is hence purely
anisotropic.

The component $\eta _{C}$ causes a type of current injection that
has not previously been noted. In all ``1+2'' experiments considered
thus far with light normally incident on a surface, the direction of
current injection lies in the plane of the surface. However, with
co-linearly polarized light fields normally incident on a $\left(
111\right) $ surface, the current can have a component into (or out
of) the surface. The current in this case is
\begin{equation}
\mathbf{\dot{J}}_{(I)} = 2\mathrm{Im} \left(
E_{\omega}^{2}E_{2\omega}^{*}\right) \left[ \bar{\eta}
\mathbf{\hat{e}} +\frac{\sqrt{2}}{6} \eta _{C}\cos \left( 3\theta
\right) \mathbf{\hat{z }} \right],
\end{equation}
where $\bar{\eta} \equiv \left( \eta _{B1}+\eta
_{B2}+\frac{1}{2}\eta _{C}\right)$, $\mathbf{\hat{z}}$ is the
$\left[ 111\right] $ direction, and $\theta$ is the angle between
$\mathbf{\hat{e}}$ and the $\left[ 2\bar{1}\bar{1}\right] $
direction. Thus, $\eta _{C}$ governs this ``surfacing'' current.

\section{Population control\label{sec:Population}}

The carrier injection rate due to the field \eqref{eq:Efield} can be
written $\dot{N}=\dot{N}_{(1)} +\dot{N}_{(I)} +\dot{N}_{(2)}$, where
$N$ is the density of electron-hole pairs, $\dot{N}_{(1)}=\xi
_{\left( 1\right) }^{ij}E_{2\omega}^{i*}E_{2\omega}^{j}$ is
one-photon absorption, $\dot{N}_{(2)}= \xi _{\left( 2\right)
}^{ijkl}E_{\omega}^{i*}E_{\omega}^{j*}E_{\omega}^{k}E_{\omega}^{l}$
is two-photon absorption, and
\begin{equation}
\dot{N}_{(I)} = \xi
_{(I)}^{ijk}E_{\omega}^{*i}E_{\omega}^{*j}E_{2\omega}^{k}+c.c.
\end{equation}
is ``1+2'' population control \cite{Fraser99PRL}. The third-rank
tensor $\xi _{(I)}^{ijk}$ has intrinsic symmetry $\xi
_{(I)}^{jik}=\xi _{(I)}^{ijk}$. In centrosymmetric materials, such
as those with the diamond structure (point group $O_{h}$), $\xi
_{(I)}^{ijk}$ is identically zero; hence, population control
requires a noncentrosymmetric material. In a material with
zinc-blende symmetry (point group $T_{d}$), $\xi _{(I)}^{ijk}$ has
only one independent component; in the standard cubic basis,
$\xi_{(I)} ^{abc}=\xi_{(I)} ^{cab}=\xi_{(I)} ^{bca}=\xi_{(I)}
^{acb}=\xi_{(I)} ^{bac}=\xi_{(I)}^{cba}$ are the only non-zero
components, where $a$, $b$, and $c$ denote components along the
principal cubic axes.

We calculate $\xi _{(I)}$ with the microscopic expression given by
Fraser et al., which was derived in the independent-particle
approximation, and is restricted to $\hbar \omega <E_{g}<2\hbar
\omega $ \cite{Fraser99PRL}. Under those conditions, $\xi _{(I)}$ is
real and is proportional to the imaginary part of the susceptibility
for second harmonic generation (SHG)
\cite{Fraser99PRL,SipeShkrebtii00}; specifically, (in mks)
\begin{equation}
\xi _{(I)}^{abc}=\frac{ 2\varepsilon _{0} }{\hbar }\mathrm{Im}\chi
^{\left( 2\right) cba}\left( -2\omega ;\omega ,\omega \right)
\label{zetaI_chi2_relation}.
\end{equation}
This connection to SHG, which can be derived from considerations of
energy transfer and macroscopic electrodynamics \cite{Fraser99PRL,
BhatSipeExcitonic05}, is important because the imaginary part of
$\chi ^{\left( 2\right) }\left( -2\omega ;\omega ,\omega \right) $
has sometimes been presented \textit{en route} to a calculation of
$\left| \chi ^{(2)}\right| $ \cite{Bell71, Moss87, Ghahramani91,
Huang93, LewYanVoon94, HughesSipe96, Adolph98}. As well, analytic
expressions have been derived for the dispersion of SHG by using
simple band models, with approximations appropriate for $2\hbar
\omega $ near the band gap \cite{Bell71, Kelley63a, Kelley63b,
Rustagi69, Bell72, Jha72}. However, these earlier works did not
connect $\mathrm{Im}\chi ^{\left( 2\right) }\left( -2\omega ;\omega
,\omega \right) $ with population control, and in fact typically
stated that it was not independently observable.

\begin{figure*}
\includegraphics[]{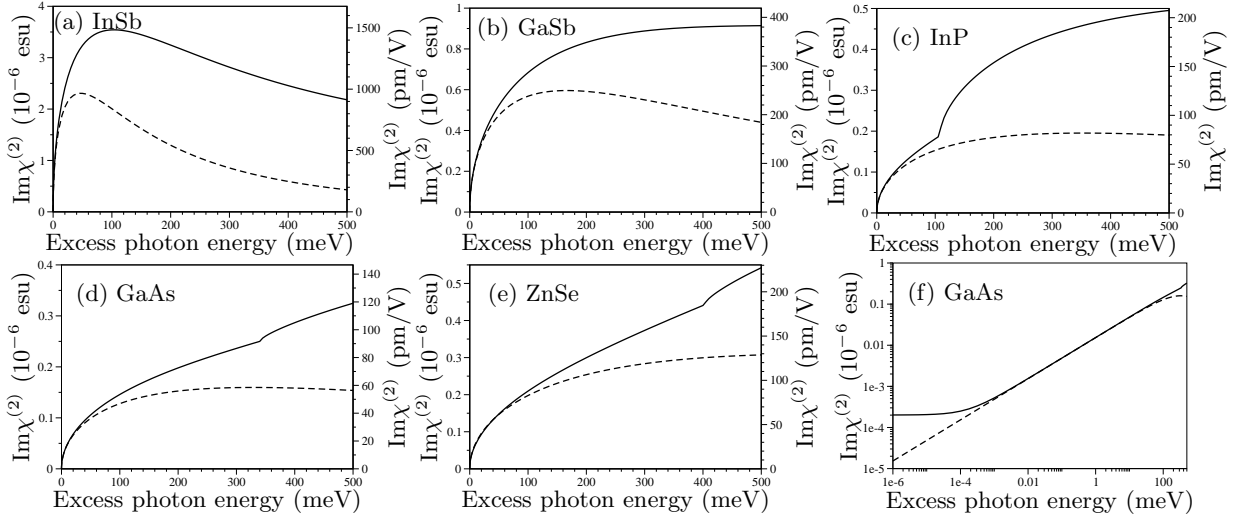}%
\caption{$\mathrm{Im}\chi ^{\left( 2\right) }$ calculated with
$H_{14}$ (solid line) and $H_{14}$-PBA (dashed line) for (a) InSb,
(b) GaSb, (c) InP, (d) GaAs, and (e) ZnSe. Panel (f) shows the GaAs
calculations on a log-log plot.}\label{fig:pop_Imchi2}
\end{figure*}
Fig.\ \ref{fig:pop_Imchi2} shows the calculation of $\mathrm{Im}\chi
^{\left( 2\right) cba}\left( -2\omega ;\omega ,\omega \right) $ for
InSb, GaSb, InP, GaAs and ZnSe. Also shown for comparison is the
PBA\ expression (\ref{eq:popControlPBA}), derived in Appendix
\ref{App:PBA_a-a}. Each spectrum can be divided into roughly three
regions. At very low excess photon energies, visible in the log-log
plot Fig.\ \ref{fig:pop_Imchi2}(f), the spectrum is roughly
independent of $\omega $. This flat part of the spectrum disappears
if $C_{k} $ is set to zero; hence, it is due to the
$\mathbf{k}$-linear term in the $c$ band spin-splitting. Next higher
in photon energy, up to about 100 meV in GaSb, InP, GaAs, and ZnSe
(up to about 15 meV in InSb), is a region where the agreement with
the analytic expression (\ref{eq:popControlPBA}) is best. In this
region, the ratio $X_{2}/X_{1}$, defined in Appendix
\ref{App:PBA_a-a}, is $0.37$ for InSb, $0.30$ for GaSb, $-0.25$ for
InP, $0.08$ for GaAs, and $0.07$ for ZnSe. At higher photon
energies, the dispersion of $\mathrm{Im}\chi ^{\left( 2\right)
cba}\left( -2\omega ;\omega ,\omega \right) $ deviates from the PBA\
expression due to band nonparabolicity and warping,
$\mathbf{k}$-dependence of matrix elements, and transitions from the
split-off band, which are not included in (\ref{eq:popControlPBA}).

If we remove the two-band transitions $hh$-$\left\{
hh,c\right\}$-$c$, $lh$-$\left\{ lh,c\right\}$-$c$, and
$so$-$\left\{ so,c\right\}$-$c$, then the calculation of
$\mathrm{Im}\chi^{\left( 2\right)}$ (or $\xi _{(I)}$) is unchanged.
This is expected for materials of zinc-blende symmetry
\cite{Kelley63b, Aspnes72}. Further, many years ago Aspnes argued
that the so-called ``virtual hole terms'' of the form
$lh$-$\{so,hh\}$-$c$ and $hh$-$\{so,lh\}$-$c$ make only a small
contribution to $\chi ^{\left( 2\right) }\left( 0\right) $
\cite{Aspnes72}. Such terms have been neglected in some previous
calculations of $\chi ^{\left( 2\right) }$ dispersion \cite{Moss87,
Huang93}. By removing the virtual hole terms, leaving only $\left\{
so,lh,hh \right\}$-$uc$-$c$ transitions, we find $\xi _{(I)}$ is
reduced by only 6--10\% over the range from the band edge to 500 meV
above the gap for GaAs. It is thus clear that inclusion of the $uc$
bands is necessary for a calculation of population control. For some
purposes it is also sufficient, since if remote band effects are
removed from the model, leaving the ``bare'' fourteen-band model
\cite{PZ90,HW94}, $\xi_{(I)}$ is decreased by only 7--10\% from its
full value for GaAs.

For most materials, the results in Fig.\ \ref{fig:pop_Imchi2} are in
reasonable agreement with previous calculations of $\mathrm{Im} \chi
^{\left( 2\right) }$ \cite{Adolph98, HughesSipe96, Huang93, Bell71},
although most previous calculations had poor spectral resolution in
this energy range. However, for ZnSe, the situation is more
complicated. The calculation of Huang and Chin is about an order of
magnitude smaller than ours \cite{Huang93}, and that of Ghahramani
et al.\ is about 5 times smaller than ours \cite{Ghahramani91}. Note
also that Huang and Chin calculated $\chi ^{\left( 2\right) }\left(
0\right) $ for ZnSe to be an order of magnitude smaller than
experimental results \cite{Huang93}. Wagner et al.\ have measured
the dispersion of $|\chi ^{\left( 2\right)}|$, which is an upper
bound on $\mathrm{Im}\chi ^{\left( 2\right)}$; for ZnSe it is about
a factor of two smaller than our calculation of $\mathrm{Im} \chi
^{\left( 2\right) }$ \cite{Wagner98}. Note that Wagner et al.\ give
a different set of band parameters than we have used here
\cite{Wagner98}.

The magnitude of $\xi _{(I)}^{abc}$ determines the magnitude of
population control, but in an experiment one is more interested in
the depth of the phase-dependent modulation of the carrier
absorption, i.e.\ the control ratio $R$ \cite{Fraser99PRL}. It is
\[
R=\frac{\dot{N}_{(I)}}{\dot{N}_{(1)}+\dot{N}_{(2)}}=\frac{\xi
_{(I)}^{ijk}E_{2\omega}^{i*}E_{\omega}^{j}E_{\omega}^{k}+c.c.}{\xi
_{(1)}^{ij}E_{2\omega}^{i*}E_{2\omega}^{j}+\xi
_{(2)}^{ijkl}E_{\omega}^{i*}E_{\omega}^{j*}E_{\omega}^{k}E_{\omega}^{l}}.
\]
This ratio is largest for field amplitudes that equalize
$\dot{N}_{(1)}$ and $\dot{N}_{(2)}$ \cite{Fraser03}; in what
follows, we assume this condition has been met. The ratio then
depends only on $\xi_{(I)}^{ijk}$, $\xi_{(1)}^{ij} $,
$\xi_{(2)}^{ijkl}$, and the polarizations of the two fields. For
light normally incident on a $\left( 111\right) $ surface,
linearly-polarized fields yield $R=\sqrt{2}\xi _{(I)}^{abc}/\sqrt{3
\xi _{(1)}^{aa} \xi _{(2)}^{aaaa} \left( 1 - \sigma /2 \right)}$,
while opposite circularly-polarized fields yield
\begin{equation}
R=2\xi _{(I)}/\sqrt{3 \xi _{(1)} \xi _{(2)}^{aaaa} \left( 1-\sigma
/6 - \delta \right) },  \label{ratio opposite circular}
\end{equation}
where $\sigma \equiv (\xi _{(2)}^{aaaa} - \xi _{(2)}^{aabb} - 2 \xi
_{(2)}^{abab})/ \xi_{(2)}^{aaaa}$ and $\delta \equiv (\xi
_{(2)}^{aaaa} + \xi _{(2)}^{aabb} - 2 \xi _{(2)}^{abab})/ (2
\xi_{(2)}^{aaaa})$ are two-photon absorption anisotropy and circular
dichroism parameters \cite{Dvorak94,HW94}. Stevens et al.\ found
that for light normally incident on a $\left( 111\right) $ surface
of GaAs, opposite circularly polarized fields yield the largest
ratio \cite{Stevens_pssb, StevensReview04}. For light normally
incident on a $\left( 110 \right)$ surface, fields linearly
polarized along $\left[ 1\bar{1}1 \right]$ yield
\begin{equation}
R=2\xi _{(I)}/\sqrt{3\xi _{(1)} \xi _{(2)}^{aaaa} \left( 1-2 \sigma
/3 \right) }.\label{e:ratio_111}
\end{equation}
The polarization configuration that yields a global maximum for the
control ratio depends on the material and photon energy; we have
found that (\ref{ratio opposite circular}) is the maximum except for
very close to the band edge, where (\ref{e:ratio_111}) is the
maximum.

To calculate the population control ratio, it is desirable to use
values of $\xi _{(I)}$, $\xi _{(1)}$, and $\xi _{(2)}$ calculated
within the same set of approximations. We use microscopic
expressions for $\xi _{(1)}$ and $\xi _{(2)}$ in the
independent-particle approximation \cite{Atanasov96}, and calculate
them within the fourteen-band model. Note that our calculation of
two-photon absorption ($\xi _{(2)}$) is similar to that of Hutchings
and Wherrett \cite{HW94}, but that our model includes remote band
effects.

\begin{figure}
\includegraphics[]{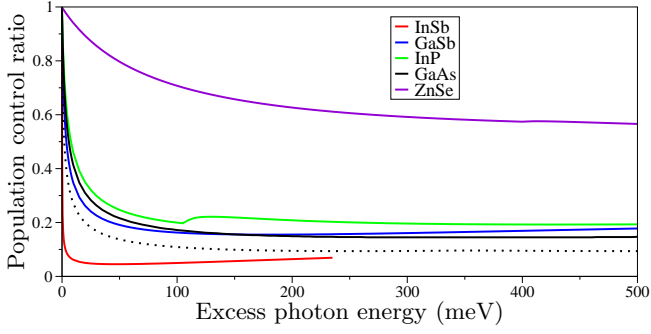}%
\caption{(color online) Calculated population control ratios
appropriate for opposite circularly polarized fields normally
incident on a $\left( 111\right) $ surface of InSb, GaSb, InP, GaAs,
and ZnSe. The blue, dotted line is the ratio for linearly polarized
fields normally incident on a $\left( 111\right) $ surface of
GaAs.}\label{fig:pop_ratio}
\end{figure}

Fig.\ \ref{fig:pop_ratio} shows the calculated spectra of the
population control ratio (\ref{ratio opposite circular}) for various
semiconductors. For each material, the ratio is close to unity at
the band edge, then drops steeply, but flattens out to some non-zero
ratio as photon energy is increased. In general, the smaller the
band gap (or conduction band effective mass) of the material, the
narrower the range over which the ratio drops, and the lower the
ratio at higher excess photon energy. Worth noting is the
particularly large ratio for ZnSe. Also plotted in Fig.\
\ref{fig:pop_ratio} is the ratio appropriate for linearly-polarized
fields normally incident on a $\left( 111\right) $ surface of GaAs,
which was the configuration in the experiment of Fraser et al
\cite{Fraser99PRL}. For all materials, the ratio (\ref{e:ratio_111})
reaches exactly unity at the band edge, in agreement with the PBA
calculation (\ref{popcontrol_ratio_bandedge}) in Appendix
\ref{App:PBA_a-a}.

The only previous theoretical calculation of the population-control
ratio, which was for GaAs, missed finding the large ratio near the
band edge because it was based on \textit{ab initio} calculations of
$\xi _{(1)}$, $\xi _{(2)}$ and $\xi _{(I)}$ that had poor spectral
resolution near the band edge \cite{Fraser99PRL}. Over the rest of
the spectrum shown in Fig.\ \ref{fig:pop_ratio}, it is about a
factor of two smaller than our calculation. This is consistent with
the previous calculation being based on a calculation of the
two-photon absorption coefficient $\xi _{(2)}$ that is too large by
comparison with other calculations \cite{HW94, Murayama95}.

The population-control ratio has been measured only in GaAs
\cite{Fraser99PRL, Fraser03, Stevens_pssb, StevensReview04,
Stevens05_110}. The measured ratios on $\left( 111\right) $-GaAs, at
excess photon energies of 180 meV \cite{Fraser99PRL, Fraser03} and
312 meV \cite{Stevens_pssb, StevensReview04} were 4 to 5 times
smaller than our calculation. Some of the difference can be
attributed to phase mismatch and large sample thickness
\cite{Fraser99PRL, Fraser03, Stevens_pssb, StevensReview04}. An
experiment on a $\left( 110\right) $-grown multiple quantum well was
complicated by an additional cascaded second harmonic effect
\cite{Stevens05_110}.\

\section{Spin current \label{sec:SpinCurrent}}

Spin-current density can be quantified by a second-rank pseudotensor
$K^{ij}$ defined as the average value of the product $v^{i}S^{j}$,
where $\mathbf{v}$ is the velocity operator and $\mathbf{S}$ is the
spin operator \cite{BhatSipe00}. Note that some authors alternately
choose the first index to represent spin and the second index to
represent velocity \cite{Rashba03}. Also, due to the spin-orbit part
of the velocity operator---the so-called ``anomalous'' velocity [the
second term in \eqref{eq:v_operator}]---$\mathbf{v}$ and
$\mathbf{S}$ do not commute, and thus $v^{i}S^{j}$ is not Hermitian.
Instead, one should take $( v^{i}S^{j} + S^{j}v^{i} ) /2$ as the
operator for spin-current. But since we neglect the anomalous
velocity (see Appendix \ref{App:kdepSO}), this is not necessary.

The spin-current injection rate due to the field \eqref{eq:Efield}
can be written
\begin{equation}
\dot{K}^{ij}= \mu _{\left( 1\right)
}^{ijkl}E_{2\omega}^{k*}E_{2\omega}^{l} + \dot{K}_{(I)}^{ij} +\mu
_{\left( 2\right)
}^{ijklmn}E_{\omega}^{k*}E_{\omega}^{l*}E_{\omega}^{m}E_{\omega}^{n},
\label{eq:phenom_K}
\end{equation}
where the pseudotensor $\mu _{\left( 1\right) }^{ijkl}$ describes
one-photon spin-current injection \cite{BhatPRL05}, the pseudotensor
$\mu _{\left( 2\right) }^{ijklmn}$ describes two-photon spin-current
injection, and
\begin{equation} \dot{K}_{(I)}^{ij} =  \mu_{(I)}
^{ijklm}E_{\omega}^{*k}E_{\omega}^{*l}E_{2\omega}^{m}+c.c.
\end{equation}
is ``1+2'' spin-current injection \cite{BhatSipe00}. The fifth-rank
pseudotensor $\mu_{(I)} ^{ijklm}$ has intrinsic symmetry $\mu_{(I)}
^{ijlkm}=\mu_{(I)} ^{ijklm}$. In an isotropic material, $\mu_{(I)}
^{ijklm}$ has three independent components, while in a cubic
material (with $T_{d}$, $O$, or $O_{h}$ symmetry) $\mu_{(I)}
^{ijklm}$ has six independent components. The four parameters
$A_{i}$, $i=1$--$4$, that we used previously to describe
spin-current injection in an isotropic model \cite{BhatSipe00} can
be reduced to three independent components with identities such as
$\varepsilon ^{ijm}\delta ^{kl}-\varepsilon ^{ijk}\delta
^{lm}+\varepsilon ^{jkm}\delta ^{il}-\varepsilon ^{ikm}\delta
^{jl}=0$ \cite{Kearsley75}. For a cubic material, $\mu_{(I)}
^{ijklm}$ has 54 non-zero elements in the principal cubic basis, and
can be written
\begin{equation}
\begin{split}
\mu_{(I)} ^{ijklm}=& \frac{\mu_{N1}}{2}\left( \varepsilon
^{jml}\delta ^{ik}+\varepsilon ^{jmk}\delta ^{il}\right)
+\mu_{N3}\varepsilon ^{ijm}\delta ^{kl} \\
& +\frac{\mu_{N2}}{2}\left( \varepsilon ^{iml}\delta
^{jk}+\varepsilon ^{imk}\delta ^{jl}\right) +\mu_{C1}\delta
^{ikln}\varepsilon ^{njm} \\ &+\mu_{C2}\delta ^{jkln}\varepsilon
^{nim}+\frac{\mu_{C3}}{2}\left( \delta ^{ijkn}\varepsilon
^{nml}+\delta ^{ijln}\varepsilon ^{nmk}\right) ,
\end{split} \label{eq:phenom_mu_form}
\end{equation}
where the non-isotropic tensor $\delta ^{ijkl}$ has nonzero
components $\delta ^{aaaa}=\delta ^{bbbb}=\delta ^{cccc}=1$, where
$a$, $b$, and $c$ denote components along the principal cubic axes.
The six independent components are $\mu_{N1}\equiv
2\mu_{(I)}^{acaba}$, $\mu_{N2}\equiv 2\mu_{(I)}^{caaba}$,
$\mu_{N3}\equiv \mu_{(I)}^{abccc}$, $\mu_{C1}\equiv
\mu_{(I)}^{abaac} -\mu_{N1}-\mu_{N3}$, $ \mu_{C2}\equiv
\mu_{(I)}^{baaac} -\mu_{N2}+\mu_{N3}$, and $\mu_{C3}\equiv
2\mu_{(I)}^{aaacb}-\mu_{N1}-\mu_{N2}$. Thus in a cubic material,
\begin{equation}
\begin{split}
\dot{K}_{(I)}^{ij}=& \mu_{N1}E_{\omega}^{*i}\left(
\mathbf{E}_{2\omega}\times \mathbf{E}_{\omega}^{*}\right)
^{j}+\mu_{N2}\left( \mathbf{E}_{2\omega}\times
\mathbf{E}_{\omega}^{*}\right)
^{i}E_{\omega}^{*j}  \\
& +\mu_{N3}\varepsilon ^{ijk}E_{2\omega}^{k}\left(
\mathbf{E}_{\omega}^{*}\cdot \mathbf{E} _{\omega}^{*}\right)
+\mu_{C3}\delta ^{ijkl}E_{\omega}^{*k}\left(
\mathbf{E}_{2\omega}\times \mathbf{E}_{\omega}^{*}\right) ^{l}
\\
& +\left( \mu_{C1}\delta ^{ikln}\varepsilon ^{njm}+\mu_{C2}\delta
^{jkln}\varepsilon ^{nim}\right)
E_{\omega}^{*k}E_{\omega}^{*l}E_{2\omega}^{m}+c.c.
\end{split}
\end{equation}
Note that the injection of $\left\langle \mathbf{v\cdot
S}\right\rangle $ is zero in a cubic material, i.e., $\dot{K}^{ij}$
is traceless. In an isotropic model, such as the one we used
previously \cite{BhatSipe00}, $\mu_{C1}=\mu_{C2}=\mu_{C3}=0$. The
connection to our previous notation is $\mu_{N1}=D\left(
A_{1}-A_{4}\right) $, $\mu_{N2}=D\left( A_{2}+A_{4}\right) $, and
$\mu_{N3}=D\left( A_{3}+A_{4}\right) $ \cite{BhatSipe00}.

The spin-current injection can be divided into a contribution from
electrons $\dot{K}^{ij}_{(I;e)}$, and a contribution from holes
$\dot{K}^{ij}_{(I;h)}$; that is,
$\dot{K}^{ij}_{(I)}=\dot{K}^{ij}_{(I;e)}+\dot{K}^{ij}_{(I;h)}$
(similarly,
$\mu^{ijklm}_{(I)}=\mu^{ijklm}_{(I;e)}+\mu^{ijklm}_{(I;h)}$).
Expressions in the PBA for both the electron and hole spin current
are given elsewhere \cite{BhatSipe00}; here we focus on the electron
spin current, since hole spin relaxation is typically very fast
\cite{Hilton02, Yu05}.

A microscopic expression for the spin-current injection was derived
previously in the Fermi's golden rule (FGR)\ limit of perturbation
theory and applied to a model in which all bands are doubly
degenerate \cite{BhatSipe00}. However, it is unsuitable for a
calculation with $H_{14}$, which accounts for the small splitting of
the spin degeneracy that occurs in materials of zinc-blende symmetry
\cite{Dresselhaus55,CCF88,PikusReview88}. If the spin-split bands
were well separated, then the microscopic expression for
$\dot{K}^{ij}_{(I;e)}$ would be
\begin{equation*}
\begin{split}
\dot{K}^{ij}_{(I;e)} =& \frac{2\pi
}{L^{3}}\sum_{c,v,\mathbf{k}}\left\langle c\mathbf{k} \right|
v^{i}S^{j}\left| c\mathbf{k}\right\rangle  \\ & \times \left[ \left(
\Omega _{c,v,\mathbf{k}}^{\left( 2\right) }\right) ^{*}\Omega
_{c,v,\mathbf{k}}^{\left( 1\right) }+c.c.\right] \delta \left(
2\omega -\omega _{cv}\left( \mathbf{k}\right) \right) ,
\end{split}
\end{equation*}
where $L^{3}$ is a normalization volume; the one-photon amplitude
$\Omega _{c,v,\mathbf{k}}^{\left( 1\right) }$ is
\begin{equation}
\Omega _{c,v,{\mathbf{k}}}^{\left( 1\right) }=i\frac{e}{2\hbar
\omega }\mathbf{E} _{2\omega } \cdot \mathbf{v}_{c,v}\left(
\mathbf{k}\right), \label{OPamplitude}
\end{equation}
where the charge on an electron is $e$ ($e<0$), and the two-photon
amplitude $\Omega _{c,v,\mathbf{k}}^{\left( 2\right) }$ is
\begin{equation}
\Omega_{c,v,\mathbf{k}}^{(2)} = \left( \frac{e}{\hbar
\omega}\right)^{2} \sum_{n}\frac{\left( \mathbf{E}_{\omega }\cdot
\mathbf{v}_{c,n}\left( \mathbf{k}\right) \right) \left(
\mathbf{E}_{\omega }\cdot \mathbf{v}_{n,v}\left( \mathbf{k}\right)
\right) }{\omega_{nv}\left( \mathbf{k}\right) -\omega  }.
\label{TPamplitude}
\end{equation}

However, for the photon energies and materials studied here, the
spin-splitting is small; it is comparable to the broadening that one
would calculate from the scattering time of the states, and also to
the laser bandwidth for typical ultrafast experiments. Thus, the
spin-split bands should be treated as quasidegenerate in FGR, with
the result
\begin{equation*}
\begin{split}
\dot{K}_{(I;e)}^{ij} =& \frac{2\pi }{L^{3}}\sum_{c,c^{\prime
}}^{\prime }\sum_{v,\mathbf{k}} \langle c\mathbf{k} | v^{i}S^{j} |
c^{\prime }\mathbf{k} \rangle \left( \Omega
_{c,v,\mathbf{k}}^{\left( 2\right) }\right) ^{*}\Omega _{c^{\prime
},v,\mathbf{k}}^{\left( 1\right) }
\\ & \times \frac{1}{2}\left[ \delta \left( 2\omega -\omega _{cv}\left(
\mathbf{k} \right) \right) +\delta \left( 2\omega -\omega
_{c^{\prime }v}\left( \mathbf{ k}\right) \right) \right] +c.c.,
\end{split}
\end{equation*}
where the prime on the summation indicates a restriction to pairs
$\left( c,c^{\prime }\right) $ for which either $c^{\prime }=c$, or
$c$ and $c^{\prime }$ are a quasidegenerate pair. The optical
excitation of the coherence between spin-split bands can be
justified using the semiconductor optical Bloch equation approach,
as was done for the one-photon spin properties \cite{BhatPRL05}.
Note that this issue does not arise for ``1+2'' current injection or
``1+2'' population control, since $\langle c \mathbf{k} | \mathbf{v}
| c^{\prime} \mathbf{k} \rangle$ and $\langle c \mathbf{k} |
c^{\prime} \mathbf{k} \rangle$ vanish between spin-split bands.

Using the time-reversal properties of the Bloch functions, we find
that $\mu_{(I;e)} $ is real, and can be written as
\begin{equation}\label{eq:mu_micro}
\begin{split}
\mu _{(I;e)}^{ijklm}=&i\left( \frac{e}{\hbar \omega }\right)
^{3}\frac{\pi }{ 2L^{3}}\sum_{c,c^{\prime }}^{\prime
}\sum_{v,\mathbf{k}}\sum_{n} \delta \left( 2\omega -\omega
_{cv}\left( \mathbf{k} \right) \right) \\ &\times \mathrm{Re}
\left\{ \frac{\left\langle c\mathbf{k}\right| v^{i}S^{j}\left|
c^{\prime }\mathbf{k}\right\rangle }{\omega _{nv\mathbf{k}}-\omega
}\left[ M_{c,c^{\prime },v}^{klm}-\left( M_{c^{\prime
},c,v}^{klm}\right) ^{*}\right] \right\} ,
\end{split}
\end{equation} where
\begin{equation}
M_{c,c^{\prime },v}^{klm}\equiv \frac{1}{2} v_{c^{\prime }v}^{m}
(\mathbf{k}) \left[ v_{cn}^{k*} (\mathbf{k}) v_{nv}^{l*}(\mathbf{k})
+ v_{cn}^{l*} (\mathbf{k}) v_{nv}^{k*} (\mathbf{k}) \right]
.\label{eq:Mccpv}
\end{equation}
That $\mu_{(I;e)}$ in \eqref{eq:mu_micro} is purely real is a
consequence of the independent-particle approximation
\cite{BhatSipeExcitonic05}.

\begin{figure}
\includegraphics[]{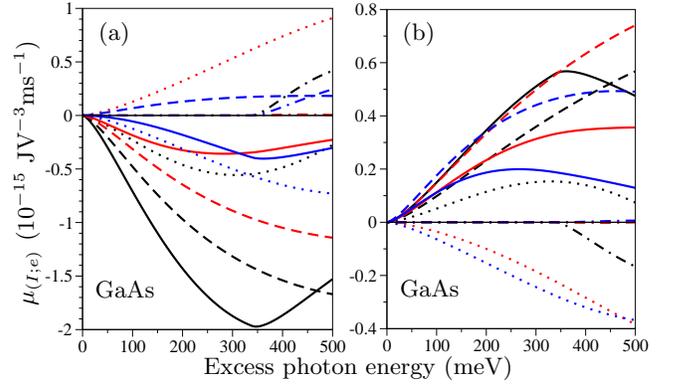}%
\caption{(color online): Calculated spectra of GaAs spin-current
injection components and their contributions from each initial
valence band; dashed, dotted, and dashed-dotted lines include only
transitions from the $hh$, $lh$, and $so$ bands respectively, while
the solid lines include all three transitions. Panel (a) shows
$\mu_{N1}$ (black lines), $\mu_{N2}$ (red lines), and $\mu_{N3}$
(blue lines). Panel (b) shows $\mu_{C1}$ (black lines), $\mu_{C2}$
(red lines), and $\mu_{C3}$ (blue
lines).}\label{fig:spincurrentGaAs_initial}
\end{figure}

\subsection{Calculation results}

\begin{figure*}
\includegraphics[]{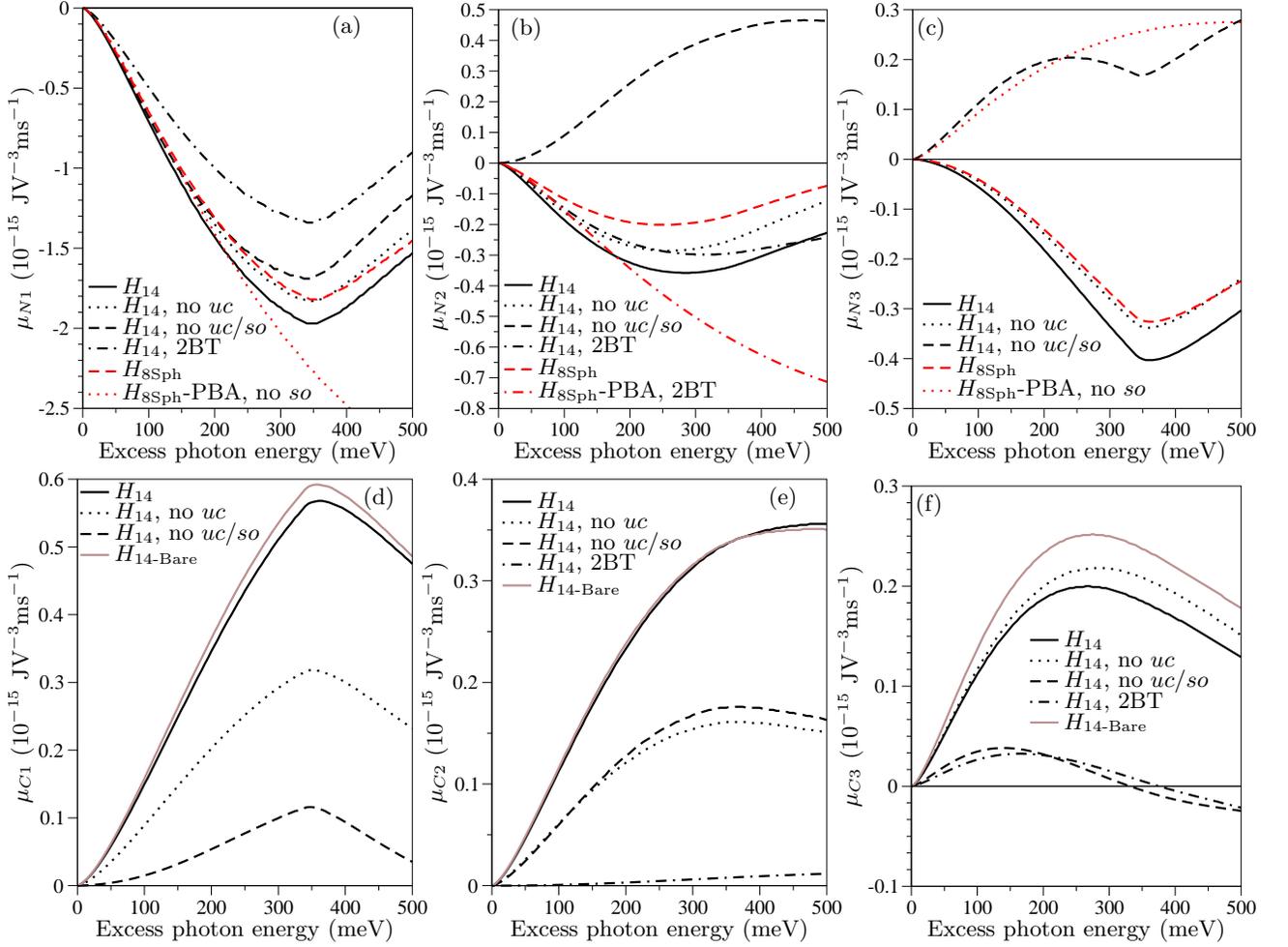}%
\caption{(color online): Approximations for GaAs spin current
components (a) $\mu_{N1}$, (b) $\mu_{N2}$, (c) $\mu_{N3}$, (d)
$\mu_{C1}$, (e) $\mu_{C2}$, (f) $\mu_{C3}$. The approximations are
described in Sec.\ \ref{approximations_defined}.}
\label{fig:spincurrentGaAs_models}
\end{figure*}

The spectra of the independent components of $\mu _{(I;e)}$,
calculated for GaAs, are shown in Fig.\
\ref{fig:spincurrentGaAs_initial} and Fig.\
\ref{fig:spincurrentGaAs_models}. Figure
\ref{fig:spincurrentGaAs_initial} also shows contributions from each
possible initial valence band. Figure
\ref{fig:spincurrentGaAs_models} shows the spin-current injection
calculated with various degrees of approximation described in Sec.\
\ref{approximations_defined}. The only other calculation of ``1+2''
spin-current injection for bulk GaAs is our earlier calculation,
which used a spherical, parabolic-band approximation to the
eight-band model and did not include the $so$ band as an
intermediate state \cite{BhatSipe00}; it is shown in Fig.\
\ref{fig:spincurrentGaAs_models} for $\mu_{N1}$, $\mu_{N2}$, and
$\mu_{N3}$.

The term $\mu_{N1}$ has the largest magnitude of the six independent
parameters of $\mu_{(I;e)}$. Since it is negative for $hh$ and $lh$
transitions but positive for $so$ transitions, it peaks in magnitude
at $2\hbar \omega $ just above $E_{g}+\Delta _{0}$ (the energy at
which $so$ transitions become allowed). Two band terms make the
largest contribution to $\mu_{N1}$, followed by three-band terms
with $hh$ or $lh$ intermediate states. The $so$ and $uc$
intermediate states make a very small contribution to $\mu_{N1}$ for
excess energies less than 200 meV. The warping of the bands is not
important for $\mu_{N1}$, since the calculation with
$H_{8\text{Sph}}$ closely approximates the ``$H_{14}$, no $uc$''
calculation, which includes the same intermediate states. The
``$H_{8\text{Sph}}$-PBA, no $so$'' calculation, which we derived
previously \cite{BhatSipe00}, is a good approximation to $\mu_{N1}$
at excess energies below 250 meV; nonparabolicity becomes important
at higher energies. The $hh$ contribution has a larger magnitude
than the $lh$ contribution in part because three-band terms increase
the magnitude of the $hh$ contribution, but decrease that of the
$lh$ contribution, as expected from the PBA expression
\eqref{eq:PBA_N1e}.

The term $\mu_{N2}$ is negative for $hh$ transitions, positive for
$lh$ transitions, and negligible for $so$ transitions. The
calculation ``$H_{14}$, 2BT'' is a good approximation to the
calculation $H_{14}$. However, the three-band terms are not small;
rather, they nearly cancel. In particular the transition
$hh$-$lh$-$c$ makes a large positive contribution to $\mu_{N2}$,
while the transition $hh$-$so$-$c$ makes a large negative
contribution. Since our earlier PBA\ calculation included the former
but not the latter \cite{BhatSipe00}, it is a poor approximation to
$\mu_{N2}$. But by including only 2BTs, it is a fair approximation
for excess energies less than 200 meV. This agreement is fortuitous,
since the calculation $H_{8\text{Sph}}$ underestimates the magnitude
of $\mu_{N2}$, and the PBA\ leads to an overestimation of the
magnitude of $\mu_{N2}$.

The term $\mu_{N3}$ is negligible when only 2BTs are included, in
agreement with the PBA \cite{BhatSipe00}. The $hh$-$lh$-$c$
transitions are positive, while the $lh$-$hh$-$c$ transitions are
negative; the former is larger, and thus $\mu_{N3}$ is positive when
$so$ intermediate states are neglected. Both $lh$-$so$-$c$ and
$hh$-$so$-$c$ are negative and substantial enough to make the total
$\mu_{N3}$ negative. Consequently, our earlier PBA result
\cite{BhatSipe00}, which neglects $so$ intermediate states, is a
poor approximation to $\mu_{N3}$. Upper conduction bands make a
fairly small contribution to $\mu_{N3}$, and warping does not seem
to be important for $\mu_{N3}$ since the calculation with
$H_{8\text{Sph}}$ is a good approximation.

As expected, the terms $\mu_{C1}$, $\mu_{C2}$, and $\mu_{C3}$ are
zero when calculated with $H_{8\text{Sph}}$.

The term $\mu_{C1}$ is negligible when only 2BTs are included.
Transitions with intermediate states in the set $\left\{
hh,lh,so\right\} $ comprise roughly two-thirds of $\mu_{C1}$. The
anisotropy of these transitions is not simply due to the warping of
the $hh$ and $lh$ bands, which we have determined by a calculation
(not shown) using $H_{8}$ without the remote band contribution to
the velocity. Rather, it comes from wave function mixing of the
$\Gamma _{8c}$ and $\Gamma _{7c}$ states into the valence and $c$
band states. The cubic anisotropy of two-photon absorption has been
attributed to such wave function mixing \cite{Dvorak94, HW94}. The
other third of the full $\mu_{C1}$ is due to transitions with the
$uc$ intermediate state, which would be forbidden close to the
$\Gamma $ point if the material were isotropic. We also note that
each three-band term makes a positive contribution to $\mu_{C1}$.

The term $\mu_{C2}$ is nearly negligible when only 2BTs are
included. Transitions from the $hh$ and $lh$ bands have opposite
sign, and those from the $so$ band are negligible. About half of
$\mu_{C2}$ is due to the transitions $hh$-$lh$-$c$ and
$lh$-$hh$-$c$, and the other half is due to transitions with the
$uc$ intermediate states. Transitions with $so$ intermediate states
are negligible. As with $\mu_{C1}$, the anisotropy of the
$hh$-$lh$-$c$ and $lh$-$hh$-$c$ transitions is due to the wave
function mixing of the $\Gamma _{8c}$ and $\Gamma _{7c}$ states into
the $hh$, $lh$, and $c$ band states.

The term $\mu_{C3}$ is positive for $hh$ transitions, negative for
$lh$ transitions, and negligible for $so$ transitions. The
transitions $hh$-$so$-$c$ and $lh$-$so$-$c$ account for most of the
value of $\mu_{C3}$, but 2BTs are not negligible. Transitions with
$uc$ intermediate states reduce the value of $\mu_{C3}$ by as much
as 10\%. Most of $\mu_{C3}$, especially at energies less than 200
meV, is due to the warping of the $hh$ and $lh$ bands. Consistent
with this, we find that remote band effects are somewhat important
for $\mu_{C3}$; when remote band effects are removed, the
calculation of $\mu_{C3}$ is about 25\% larger than the full
calculation. Note that $\mu_{C3}$ is far more sensitive to remote
band effects than any other optical property calculated in this
article.

\begin{figure*}
\includegraphics[]{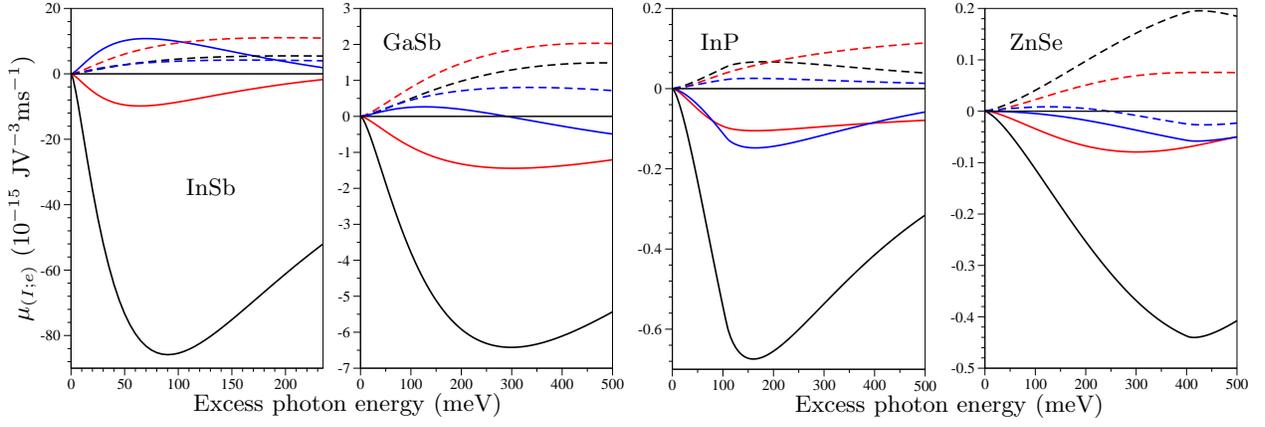}%
\caption{Spectra of spin current components for InSb, GaSb, InP, and
ZnSe: $\mu_{N1}$ (solid black line), $\mu_{N2}$ (solid red line),
$\mu_{N3}$ (solid blue line), $\mu_{C1}$ (dashed black line),
$\mu_{C2}$ (dashed red line), and $\mu_{C3}$ (dashed blue line).}
\label{fig:spincurrent_4materials}
\end{figure*}

In Fig.\ \ref{fig:spincurrent_4materials} we plot the spectra of the
independent components of the spin current density pseudotensor for
InSb, GaSb, InP, and ZnSe. The spin current tensor is largest for
InSb in agreement with the PBA expressions in Appendix
\ref{App:PBA_a-a}. We also note that $\mu_{N3}$ is positive for InSb
and GaSb at low excess photon energy, whereas it is negative for
InP, GaAs, and ZnSe.

\subsection{Configurations}
\begin{figure}
\includegraphics[]{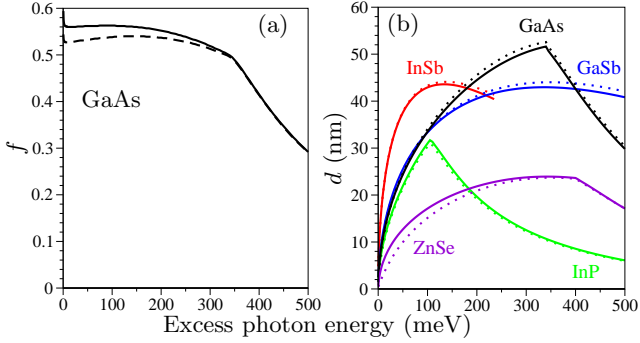}%
\caption{(a) Degree of polarization of spin-polarized current due to
co-circularly polarized fields. (b) Displacement of spins in pure
spin current due to cross-linearly polarized
fields.}\label{fig:SC_measures}
\end{figure}

Co-circularly polarized fields generate a spin-polarized current,
which can be characterized by its degree of spin polarization
$f\equiv \left(2e/\hbar \right)
\dot{K_{e}^{ij}}\hat{q}^{i}\hat{n}^{j}/|\dot{\mathbf{J}_{e}}|$,
where $\hat{\mathbf{n}}$ is a unit vector normal to the polarization
plane of the fields, and $\hat{\mathbf{q}}$ is a unit vector in the
direction of $\mathbf{J}_{e}$ \cite{BhatSipe00}. Essentially,
$f=\left\langle vS\right\rangle / \left\langle v\right\rangle$.
Since this measure aims to characterize the photoexcited
distribution of electrons, we neglect holes from both $\dot{K}$ and
$\dot{J}$ in this calculation \cite{footnote2000f}. For fields
normally incident on a $\left( 001\right) $ surface (i.e.\
$\mathbf{e}_{\omega}=\mathbf{e}_{2\omega}=\left( \mathbf{\hat{x}}\pm
i\mathbf{\hat{y}}\right) /\sqrt{2}$), the spin current is
\begin{equation*}
\begin{split}
\dot{K}_{(I)}^{ij} =& \mp \sqrt{2}|E_{2\omega}| |E_{\omega}|^{2}  \\
& \times \left[ \left( \mu_{N1}+ \frac{\mu_{C1}}{2}\right)
\hat{m}_{\pm }^{i}\hat{z}^{j}+\left(
\mu_{N2}+\frac{\mu_{C2}}{2}\right) \hat{z}^{i}\hat{m}_{\pm
}^{j}\right],
\end{split}
\end{equation*}
where $\mathbf{\hat{m}}_{\pm }=\sin \left( 2\phi _{\omega}-\phi
_{2\omega}\right) \mathbf{\hat{x}}\pm \cos \left( 2\phi
_{\omega}-\phi _{2\omega}\right) \mathbf{\hat{y}}$, the current is
$\dot{\mathbf{J}}_{(I)} = \sqrt{2} E_{2\omega} E_{\omega}^{2} \left(
\eta _{B1}+\eta _{C} /2 \right) \hat{\mathbf{m}}_{\pm }$, and the
degree of spin polarization is
\begin{equation}
f=\frac{2e}{\hbar }\frac{\mu_{N1}+\mu_{C1}/2}{\eta _{B1}+\eta
_{C}/2}.\label{eq:SC:f:001}
\end{equation}
For fields normally incident on a $\left( 111\right) $ surface,
$\dot{\mathbf{J}}_{(I)}=\sqrt{2}E_{2\omega}E_{\omega}^{2}\left( \eta
_{B1}+\eta _{C}/3\right) \hat{\mathbf{m}}_{\pm }$, and
\begin{equation}
f=\frac{2e}{\hbar }\frac{\mu_{N1}+\mu_{C1}/3+\mu_{C3}/3}{\eta
_{B1}+\eta _{C}/3}.\label{eq:SC:f:111}
\end{equation}
The degree of spin polarization is plotted for GaAs in Fig.\
\ref{fig:SC_measures}(a). The cubic anisotropy is small, but clearly
seen, especially at low excess photon energies. The other materials
have very similar degrees of spin polarization.

A pure spin current, without an electrical current, can be generated
with cross-linearly polarized fields \cite{BhatSipe00}. We consider
fields polarized in the $\left( 001\right) $ plane, with the $\omega
$ field polarized at an angle $\theta $ to the $\mathbf{\hat{x}}$
axis (i.e.\ $\left[ 100\right] $) and the $2\omega $ field polarized
at an angle $\theta $ to the $\mathbf{\hat{y}}$ axis
($\mathbf{e}_{\omega} = \mathbf{\hat{x}} \cos \theta +
\mathbf{\hat{y}} \sin \theta$ and $\mathbf{e}_{2\omega} = -
\mathbf{\hat{x}} \sin \theta + \mathbf{\hat{y}} \cos \theta$). The
spin current is
\begin{widetext}
\begin{equation*}
\begin{split}
\dot{K}_{(I)}^{ij} =& -\frac{1}{2} |E_{2\omega}|
|E_{\omega}|^{2}\cos \left( 2\phi _{\omega}-\phi _{2\omega}\right)
\left[ \left( 4\mu_{N1}+4\mu_{N3} +3\mu_{C1}+\mu_{C1}\cos \left(
4\theta \right) \right) e_{\omega}^{i}\hat{z}^{j} \right.\\ &
\left.-\sin \left( 4\theta \right)\left(\mu_{C1} e_{2\omega}^{i}
\hat{z}^{j} +\mu_{C2} \hat{z}^{i}e_{2\omega}^{j}\right) +\left(
4\mu_{N2}-4\mu_{N3}+3\mu_{C2}+\mu_{C2}\cos \left( 4\theta \right)
\right) \hat{z} ^{i}e_{\omega}^{j} \right]
\end{split}
\end{equation*}
\end{widetext}
This pure spin current is typically measured by the resulting
displacement of up and down spins \cite{StevensPRL03, HubnerPRL03}.
The finite displacement results from transport and scattering of the
electrons. Using the Boltzmann transport equation in the relaxation
time approximation with space-charge effects justifiably neglected
\cite{HubnerPRL03}, one finds $d^{i}\left( \mathbf{\hat{z}} \right)
=\left( 4\tau /\hbar \right) \dot{K}^{ij}\hat{z}^{j}/\left( \dot{N}
_{\left( 1\right) }+\dot{N}_{\left( 2\right) }\right) $
\cite{BhatPRL05}. Here, $\mathbf{d}\left( \mathbf{\hat{z}}\right) $
is the displacement of spins measured with respect to the
quantization direction $\mathbf{\hat{z}}$, and $\tau $ is the
momentum relaxation time. We assume the field intensities have been
chosen to balance one- and two-photon absorption, a condition that
is $\theta $-dependent due to the cubic anisotropy of two-photon
absorption. Thus,
\begin{equation}
\mathbf{d}\left( \mathbf{\hat{z}}\right) \cdot
\mathbf{e}_{\omega}=\frac{\tau}{\hbar} \frac{\left(
4\mu_{N1}+4\mu_{N3}+3\mu_{C1}+\mu_{C1}\cos \left( 4\theta \right)
\right) }{\sqrt{\xi_{(1)} ^{xx}\xi _{(2)}^{xxxx}\left(
1-\left(\sigma/2\right) \sin^{2} \left( 2\theta \right) \right)
}}\label{eq:SC:separationdistance_w}
\end{equation}
and
\begin{equation}
\mathbf{d}\left( \mathbf{\hat{z}}\right) \cdot
\mathbf{e}_{2\omega}=\frac{\tau}{\hbar} \frac{\mu_{C1}\sin \left(
4\theta \right) }{\sqrt{\xi_{(1)} ^{xx}\xi _{(2)}^{xxxx}\left(
1-\left(\sigma/2\right) \sin^{2} \left( 2\theta \right) \right)
}},\label{eq:SC:separationdistance_2w}
\end{equation}
where $\sigma $ is the two-photon absorption cubic-anisotropy factor
given explicitly in the next section \cite{Dvorak94, HW94}. At
$\theta =0$ and $\theta =\pi /4$, $\mathbf{d}$ is parallel to
$\mathbf{e}_{\omega}$. The spin separation distance is plotted in
Fig.\ \ref{fig:SC_measures}(b), where we have assumed a momentum
relaxation time of 100 fs for each material.

This calculation of the spin separation distance is a significant
improvement over our initial calculations \cite{StevensPRL03,
HubnerPRL03}, which used the eight-band PBA and neglected three-band
terms from the two-photon amplitude (``$H_{8\text{Sph}}$-PBA,
2BT''). Stevens et al.\ measured a spin separation distance of 20 nm
in a GaAs multiple quantum well at an excess photon energy of 200
meV, and estimated a momentum relaxation time of $\tau=45$ fs
\cite{StevensPRL03}. For $\tau=45$ fs, we calculate a spin
separation distance of 20.0 nm for bulk GaAs at 200 meV. H\"{u}bner
et al.\ measured a spin separation distance of 24 nm (the
photoluminescence spot separation is half this distance) in cubic
ZnSe at an excess photon energy of 280 meV, and estimated a momentum
relaxation time of $\tau=100$ fs \cite{HubnerPRL03}. The calculation
in Fig.\ \ref{fig:SC_measures}(b) yields $d=23.6$ nm for ZnSe at 280
meV. In both cases, we now find very good agreement with the
experiment, whereas the previous model resulted in larger spin
separation distances. Of course, this agreement is contingent on the
accuracy of the momentum relaxation time estimates.

Note that both the degree of spin polarization for co-circularly
polarized fields and the spin-separation distance, plotted in Fig.\
\ref{fig:SC_measures}, have a kink at excess photon energy
$\Delta_{0}$ and decrease at higher excess photon energies. A
similar kink and decrease, due to the onset of transitions from the
split-off band, occurs for both one-photon spin injection
\cite{DyakonovOptOrient} and two-photon spin injection
\cite{Bhat_TPS_05}.\

\section{Spin control\label{sec:Spin}}

The spin injection rate due to the field \eqref{eq:Efield} can be
written $\dot{\mathbf{S}} = \dot{\mathbf{S}}_{(1)} +
\dot{\mathbf{S}}_{(I)} + \dot{\mathbf{S}}_{(2)}$, where $\mathbf{S}$
is the macroscopic spin density, $\dot{S}_{(1)}^{i}= \zeta _{\left(
1\right) }^{ijk}E_{2\omega}^{j*}E_{2\omega}^{k}$ is one-photon spin
injection \cite{DyakonovOptOrient}, $\dot{S}_{(2)}^{i} = \zeta
_{(2)}^{ijklm}E_{\omega}^{j*}E_{\omega}^{k*}E_{\omega}^{l}E_{\omega}^{m}$
is two-photon spin injection \cite{Bhat_TPS_05}, and
\begin{equation}
\dot{S}_{(I)}^{i}= \zeta_{(I)}^{ijkl} E_{\omega}^{*j}
E_{\omega}^{*k} E_{2\omega}^{l}+c.c.
\label{eq:macro:spin}
\end{equation}
is ``1+2'' spin control \cite{Stevens_pssb}. In previous sections
and in some of the expressions below, we use $\mathbf{S}$ to denote
the single-particle spin operator. It should be obvious by context
when $\mathbf{S}$ refers to the macroscopic spin density and when it
refers to that spin operator.

The fourth-rank pseudotensor $\zeta _{(I)}^{ijkl}$ has intrinsic
symmetry on the indices $j\leftrightarrow k$. Such a pseudotensor is
zero in the presence of inversion symmetry; hence, ``1+2'' spin
control requires materials of lower symmetry. For zinc-blende
symmetry (point group $T_{d}$), a general fourth-rank pseudotensor
has three independent parameters and 18 non-zero elements in the
standard cubic basis; forcing the $j\leftrightarrow k$ symmetry
leaves two independent parameters
\begin{subequations}
\begin{align}
i\zeta _{IA} & \equiv \zeta _{(I)}^{abba}=\zeta _{(I)}^{caac}=\zeta
_{(I)}^{bccb}=-\zeta _{(I)}^{acca}=-\zeta _{(I)}^{cbbc}=-\zeta
_{(I)}^{baab} , \label{macro_zetaIA}
\\
\begin{split}
i\zeta _{IB} & \equiv \zeta _{(I)}^{aabb}=\zeta _{(I)}^{ccaa}=\zeta
_{(I)}^{bbcc}=-\zeta _{(I)}^{aacc}=-\zeta _{(I)}^{ccbb}=-\zeta _{(I)}^{bbaa} \\
& =\zeta _{(I)}^{abab}=\zeta _{(I)}^{caca}=\zeta
_{(I)}^{bcbc}=-\zeta _{(I)}^{acac}=-\zeta _{(I)}^{cbcb}=-\zeta
_{(I)}^{baba} . \label{macro_zetaIB}
\end{split}
\end{align}
\end{subequations}

The spin injection has a contribution from electrons
$\mathbf{\dot{S}}_{(I;e)}$, and a contribution from holes
$\mathbf{\dot{S}}_{(I;h)}$; that is,
$\mathbf{\dot{S}}_{(I)}=\mathbf{\dot{S}}_{(I;e)}+\mathbf{\dot{S}}_{(I;h)}$,
and $\zeta_{(I)}=\zeta_{(I;e)} + \zeta_{(I;h)}$.

We treat the spin-split bands as quasidegenerate when taking the FGR
limit of perturbation theory, as discussed for the spin current in
Section \ref{sec:SpinCurrent}, deriving the microscopic expression
\begin{equation*}
\begin{split}
\mathbf{\dot{S}}_{(I;e)}=&\frac{2\pi }{L^{3}}\sum_{c,c^{\prime
}}^{\prime }\sum_{v,\mathbf{k}}\left\langle c\mathbf{k}\right|
\mathbf{S}\left| c^{\prime }\mathbf{k}\right\rangle \left( \Omega
_{c,v,\mathbf{k}}^{\left( 2\right) }\right) ^{*}\Omega _{c^{\prime
},v,\mathbf{k}}^{\left( 1\right) } \\ & \times \frac{1}{2}\left[
\delta \left( 2\omega -\omega _{cv}\left( \mathbf{k} \right) \right)
+\delta \left( 2\omega -\omega _{c^{\prime }v}\left( \mathbf{
k}\right) \right) \right] +c.c.,
\end{split}
\end{equation*}
where the prime on the summation indicates a restriction to pairs
$\left( c,c^{\prime }\right) $ for which either $c^{\prime }=c$, or
$c$ and $c^{\prime }$ are a quasidegenerate pair. Using the
time-reversal properties of the Bloch functions, we find that
$\zeta_{(I;e)} $ is purely imaginary and can be written
\begin{equation}
\begin{split}
\zeta _{(I;e)}^{ijkl}=& i\left( \frac{e}{\hbar \omega }\right)
^{3}\frac{\pi }{ 2L^{3}}\sum_{c,c^{\prime }}^{\prime
}\sum_{v,\mathbf{k}}\sum_{n} \delta \left( 2\omega -\omega
_{cv}\left( \mathbf{k} \right) \right) \\ & \times \mathrm{Re}
\left\{ \frac{\left\langle c\mathbf{k}\right| S^{i}\left| c^{\prime
} \mathbf{k}\right\rangle }{\omega _{nv\mathbf{k}}-\omega }\left[
M_{c,c^{\prime },v}^{jkl}+\left( M_{c^{\prime },c,v}^{jkl}\right)
^{*}\right] \right\} ,\label{eq:zeta_micro}
\end{split}
\end{equation}
where $M_{c,c^{\prime },v}^{jkl}$ is given in Eq.\ (\ref{eq:Mccpv}).

\begin{figure}
\includegraphics[]{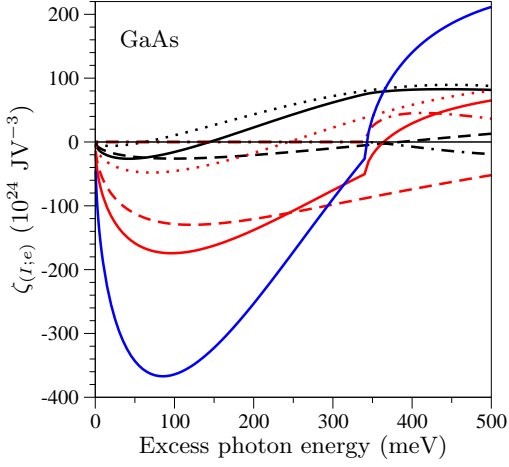}%
\caption{(color online) Spin control pseudotensor components
$\zeta_{IA}$ (black lines), $\zeta_{IB}$ (red lines), and $\left(
\zeta _{IA}+2\zeta _{IB} \right)$ (blue line) with breakdown into
initial states. Dotted lines include transitions from the $lh$ band,
dashed lines include transitions from the $hh$ band, dashed-dotted
lines include transitions from the $so$ band, and solid lines
include all transitions.} \label{fig:spin_GaAs_initials}
\end{figure}
\begin{figure}
\includegraphics[]{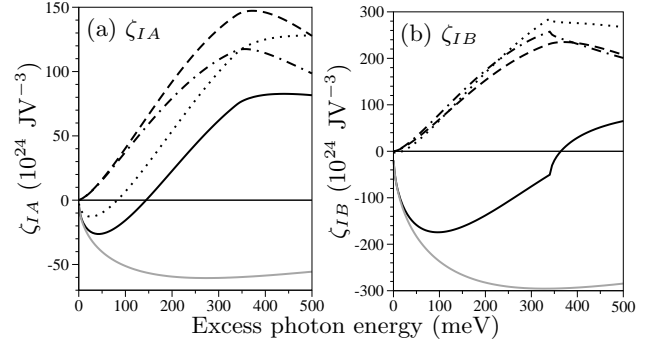}%
\caption{Spin control pseudotensor for GaAs. The calculations are
$H_{14}$ (solid black lines), ``$H_{14}$, no $uc$'' (dotted lines),
``$H_{14}$, no $uc$/$so$'' (dashed-dotted lines), ``$H_{14}$, 2BT''
(dashed lines), and ``$H_{14}$-PBA'' (solid grey lines).}
\label{fig:spin_GaAs_models}
\end{figure}

The spectra of $\zeta _{IA}$ and $\zeta _{IB}$ for GaAs are shown in
Figs.\ \ref{fig:spin_GaAs_initials} and \ref{fig:spin_GaAs_models}.
Figure \ref{fig:spin_GaAs_initials} also shows the contributions
from each possible initial valence band. Figure
\ref{fig:spin_GaAs_models} shows the spin control calculated with
various degrees of approximation described in Sec.\
\ref{approximations_defined}.

The term $\zeta _{IA}$ decreases from zero at the band edge to a
maximum negative value at 40 meV, mostly due to transitions from the
$hh$ band, and is positive at higher excess photon energies, mostly
due to transitions from the $lh$ band. The low energy behavior is in
agreement with the PBA result (\ref{eq:spin_IA_PBA}), in which the
ratio of $hh:lh$ transitions is $\left( m_{c,hh}/m_{c,lh}
\right)^{3/2}$. Transitions with $so$ and $uc$ intermediate states
dominate the decrease in $\zeta_{IA}$ at low excess photon energies,
as seen in Fig.\ \ref{fig:spin_GaAs_models}(a); they are the only
non-zero transitions in the PBA result (\ref{eq:spin_IA_PBA}). The
contribution from $uc$ intermediate states is negative and
approximately constant over most of the spectrum, whereas the
contribution from $so$ intermediate states changes from negative to
positive as transitions from the $so$ band become allowed ($2\hbar
\omega >E_{g}+\Delta_{0}$). The contribution from 2BTs, which is
zero in the PBA, is positive over the whole spectrum. The breakdown
of the PBA is due to the increase in magnitude of the 2BTs. In fact,
the sum of the PBA and the 2BTs is a good approximation to the full
calculation. We also note that a calculation with $H_{8}$ for $\zeta
_{IA}$ yields a nearly negligible result; thus, the contribution
from intermediate states within the set $\left\{ so,lh,hh,c\right\}
$ (including 2BTs) is due to the mixing of the $\Gamma _{7c}$ and
$\Gamma _{8c}$ wavefunctions with these states.

The term $\zeta _{IB}$ is larger in magnitude than the term $\zeta
_{IA}$ over most of the calculated spectrum. It falls to a maximum
negative value at 95 meV, sharply increases when transitions from
the $so$ band become allowed, and is positive at higher excess
photon energy. At lower photon energies, transitions from the $hh$
band and transitions from the $lh$ band both make negative
contributions to $\zeta _{IB}$; in the PBA result
(\ref{eq:spin_IB_PBA}) the ratio of $hh:lh$ transitions is $\left(
m_{c,hh}/m_{c,lh} \right)^{3/2}$. Fig.\
\ref{fig:spin_GaAs_models}(b) reveals that $\zeta _{IB}$ is
essentially due to contributions from $uc$ intermediate states, and
2BTs. Over the whole spectrum, the former are negative while the
latter are positive. The smallness of the contribution from $so$
intermediate states is also seen in the PBA result
(\ref{eq:spin_IB_PBA}), since $Z_{+} \gg Z_{-}^{\prime}$ in that
expression. We also note that a calculation with $H_{8}$ for $\zeta
_{IB}$ yields a nearly negligible; thus, the contribution from
intermediate states within the set $\left\{ so,lh,hh,c\right\} $
(including 2BTs) is due to the mixing of the $\Gamma _{7c}$ and
$\Gamma _{8c}$ wave functions with these states.

\begin{figure*}
\includegraphics[]{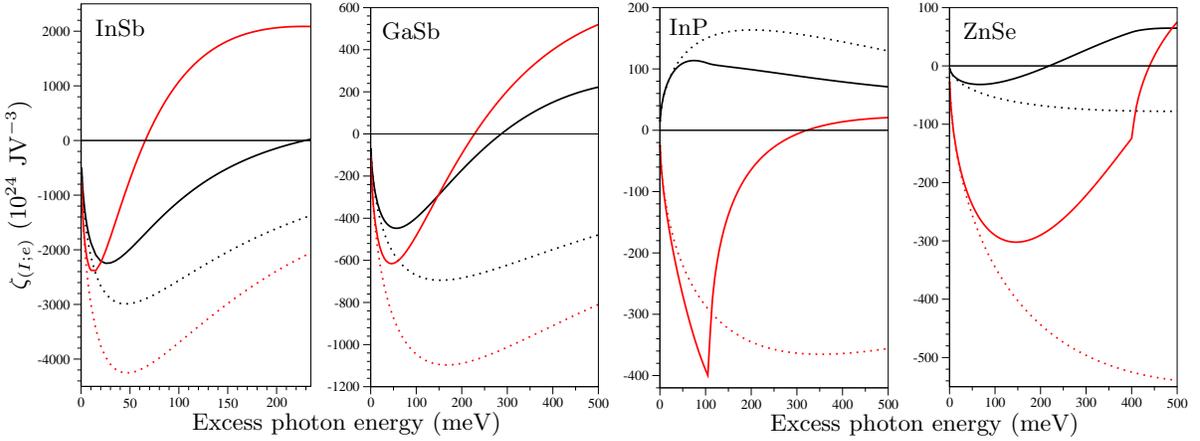}%
\caption{(color online) Spin control calculated for InSb, GaSb, InP,
and ZnSe. Black lines are $\zeta_{IA}$ and red lines are
$\zeta_{IB}$. Solid lines are the full calculation using $H_{14}$.
Dotted lines are the parabolic-band approximation calculated with
(\ref{eq:spin_IA_PBA}) and (\ref{eq:spin_IB_PBA}).}
\label{fig:spin_4materials}
\end{figure*}

We have also calculated the spin-control pseudotensor for the
semiconductors InSb, GaSb, InP, and ZnSe. The results are shown in
Fig.\ \ref{fig:spin_4materials} along with the parabolic-band
approximations (\ref{eq:spin_IA_PBA}) and (\ref{eq:spin_IB_PBA}).

The magnitude of spin control is determined by $\zeta_{(I)}$, but in
an experiment one is more interested in the depth of the
phase-dependent modulation of the spin polarization signal. One
possible definition for the signal is the ratio of spin injection
measured with both $\omega$ and $2\omega$ fields to the sum of the
spin injections measured with circularly polarized fields of each
frequency \cite{Stevens_pssb}. The amplitude of its modulation is
\begin{equation}
\frac{\left| \dot{S}_{(I)}^{z} \right|}{\dot{S}_{(1)}^{z}\left(
\sigma ^{+}\right) +\dot{S}_{(2)}^{z}\left( \sigma ^{+}\right) },
\label{eq:spin_ratio_spin_norm}
\end{equation}
where the argument $\left( \sigma ^{+}\right)$ indicates injection
with a $\sigma ^{+}$ polarized field. This ratio, which is largest
for field amplitudes that equalize $\dot{S}_{(1)}^{z}\left( \sigma
^{+}\right)$ and $\dot{S}_{(2)}^{z}\left( \sigma ^{+}\right)$, was
measured by Stevens et al.\ with excess photon energies of 150 meV
and 280 meV \cite{Stevens_pssb, Stevens05_110}.

The ratio \eqref{eq:spin_ratio_spin_norm} has an undesirable
feature: it can exceed unity. Close to the band edge in many
semiconductors (at 2 meV in GaAs), there is a photon energy for
which $\dot{S}_{(2)}^{z}\left( \sigma ^{+}\right) =0$
\cite{Bhat_TPS_05}. At that photon energy, it is impossible to
choose field amplitudes to balance one- and two-photon spin
injection with circular polarized fields [$\dot{S} _{(1)}^{z}\left(
\sigma ^{+}\right) =\dot{S}_{(2)}^{z}\left( \sigma ^{+}\right) $],
and thus the maximum ratio has a singularity. Even if the condition
$\dot{S}_{(1)}\left( \sigma ^{+}\right) = \dot{S}_{(2)}\left( \sigma
^{+}\right) $ is relaxed, the ratio \eqref{eq:spin_ratio_spin_norm}
can exceed unity. This is because $\dot{S} _{(1)}^{z}\left( \sigma
^{+}\right) $ and $\dot{S}_{(2)}^{z}\left( \sigma ^{+}\right) $ have
opposite sign close to the band gap \cite{Bhat_TPS_05}, and thus it
is possible, by appropriate choice of field amplitudes, to make the
denominator of the ratio arbitrarily small.

An alternate ratio to characterize the spin control, which has an
upper bound of unity, is
\begin{equation}
R_{S}=\frac{2}{\hbar }\frac{\left| \dot{S}_{(I)}^{z}
\right|}{\dot{N}_{(1)}+\dot{N}_{(2)}} .
\label{eq:spin_ratio_population_norm}
\end{equation}
It is the amplitude of phase-dependent oscillation of the degree of
spin polarization, and it is most useful when there is little or no
population control. We assume the fields are chosen to balance one-
and two-photon absorption. For most photon energies and materials
this is nearly the same as balancing one- and two-photon spin
injection.

For normal incidence on a $\left( 111\right) $ sample, opposite
circularly polarized fields yield
\[
R_{S}=\frac{2}{\hbar }\frac{\left| \zeta _{IA}+2\zeta _{IB} \right|}
{\sqrt{3\xi _{(1)}^{aa}\xi _{(2)}^{aaaa}\left( 1-\sigma /6-\delta
\right) }} .
\]
For normal incidence on a $\left( 110\right) $ sample, opposite
circularly polarized fields
\[
R_{S}=\frac{2}{\hbar }\frac{3\left| \zeta _{IA}+2\zeta _{IB}\right|
}{4\sqrt{ 2\xi _{(1)}^{aa}\xi _{(2)}^{aaaa}\left( 1-\delta -\sigma
/8\right) }},
\]
and orthogonal linearly polarized fields ($xy$-polarized) yield
\[
R_{S}\left( \alpha \right) =\frac{2}{\hbar }\frac{\left| \left(
\zeta _{IA}+2\zeta _{IB}\right) \left( r+3\sin ^{2}\alpha \right)
\cos \alpha \right|} {2\sqrt{\xi _{(1)}^{aa}\xi _{(2)}^{aaaa}\left(
1-\frac{1}{2}\sigma \left( \sin ^{2}\alpha \right) \left( 1+3\cos
^{2}\alpha \right) \right) }},
\]
where $r\equiv -2 \zeta_{IA} / (\zeta_{IA} + 2 \zeta_{IB})$
\cite{Stevens_pssb}, and $\alpha$ is the angle between the
polarization of the $\omega$ field ($\mathbf{E}_{\omega}$) and the
$\left[ 001\right] $ axis, which lies in the $\left( 110\right) $
plane. The angle that maximizes $R_{S}$ depends on photon energy
through $r$ and $\sigma $. We determine it numerically.

\begin{figure}
\includegraphics[]{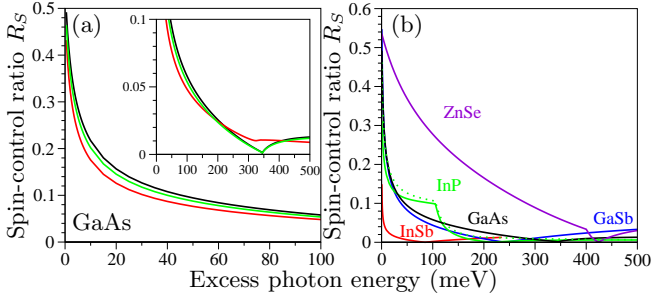}%
\caption{(color online) Spin-control ratio normalized by carrier
population [Eq.\ (\ref{eq:spin_ratio_population_norm})]. In (a), for
GaAs, black lines are $\left( 111\right)$-incident, opposite
circularly polarized fields; green lines are $\left(
110\right)$-incident, opposite circularly polarized fields; and red
lines are $\left( 110\right)$-incident, orthogonal linearly
polarized fields. In (b), for InSb, GaSb, InP, GaAs, and ZnSe, solid
lines are $\left( 111\right) $-incident, opposite circularly
polarized fields. The dotted line in (b) is $\left( 110\right)
$-incident, orthogonal linearly polarized fields for InP.}
\label{fig:spin_ratio}
\end{figure}
The ratio $R_{S}$ for GaAs is plotted in Fig.\
\ref{fig:spin_ratio}(a). For $\left( 111\right)$-incidence, opposite
circularly polarized fields yield the highest ratio over the studied
range of photon energies. For $\left( 110\right) $-incidence,
opposite circularly polarized fields yield the highest ratio, except
for between 190 meV and 415 meV when $xy$-polarized fields the
highest ratio. For $xy$-polarized fields, the angle that yields the
largest ratio decreases from 0.99 rad to 0.53 rad from the band edge
to 320 meV, and is zero for higher excess energies.

The ratio $R_{S}$ for the five semiconductors InSb, GaSb, InP, GaAs,
and ZnSe are plotted in Fig.\ \ref{fig:spin_ratio}(b). At low photon
energy, opposite circularly polarized fields normally incident on
$\left( 111\right) $ yield the largest ratio for InSb, GaSb, GaAs,
and ZnSe, whereas orthogonal linearly polarized fields normally
incident on $\left(110\right)$ yield the largest ratio for InP.

\section{Summary\label{sec:Summary}}

We have studied the four ``1+2'' coherent control effects---current
injection, spin-current injection, population control, and spin
control---in bulk semiconductors having zinc-blende symmetry. We
used an empirical, fourteen-band $\mathbf{k} \cdot \mathbf{p}$
Hamiltonian and examined the relative importance to each effect of
the possible initial and intermediate states. We have also studied
the crystal orientation and polarization dependencies of each
effect. Cubic anisotropy is small in some cases, but large in
others.

We have compared the numerical calculation with analytic
expressions, derived in the parabolic-band approximation, to show
the value and limitations of the latter. The PBA expressions, where
they are accurate, are useful to show how the effects scale in
different materials.

The comparison between the two approaches establishes that, at low
excess photon energies, ``1+2'' current injection and ``1+2''
spin-current injection are due to interference of allowed one-photon
transitions and allowed-forbidden two-photon transitions, whereas
``1+2'' population control and ``1+2'' spin control are due to
interference of allowed one-photon transitions and allowed-allowed
two photon transitions. It also explains the large population- and
spin-control ratios predicted by the fourteen-band calculation close
to the band edge, where allowed-allowed two-photon transitions
dominate allowed-forbidden two-photon transitions. Neither ``1+2''
population control, nor ``1+2'' spin control have yet been
experimentally studied in that spectral range.

\appendix
\section{Neglect of the anomalous velocity and $\mathbf{k}$-dependent spin-orbit
coupling\label{App:kdepSO}}

The anomalous velocity, i.e.\ $\mathbf{v}_{A}\equiv \left(
\mathbf{v}-\mathbf{p}/m\right) =\hbar \left( \bm{\sigma }\times
\bm{\nabla } V\right) /\left( 4m^{2}c^{2}\right) $, which leads to
$\mathbf{k}$-dependent spin-orbit coupling in $H_{\mathbf{k}}$ from
the term $\hbar \mathbf{k}\cdot \mathbf{v}_{A}$, is often neglected
in $\mathbf{k}\cdot \mathbf{p}$ models \cite{Kane56,
CardonaPollak66, Rossler84, PZ90, PZ96}. Some authors have treated
matrix elements of $\bm{\nabla }V$ as additional independent
parameters \cite{Dresselhaus55, Rustagi69, Bahder90, Ostromek96}.
For example, Bahder, who gives the matrix for $\hbar \mathbf{k}\cdot
\mathbf{v}_{A}$ within the eight-band model, defines the model
parameter \cite{Bahder90, footnoteC0Ck}
\[
C_{0} \equiv \frac{1}{\sqrt{3}}\frac{\hbar
^{2}}{4m^{2}c^{2}}\left\langle S\right| \nabla _{x}V\left|
X\right\rangle .
\]
Ostromek used the value $C_{0}=0.16$ eV {\AA} to fit the eight-band
model to experimental results \cite{Ostromek96}. We here relate
matrix elements of $\bm{\nabla }V$ (and hence matrix elements of
$\mathbf{v}_{A}$) to other parameters of the model, thereby
demonstrating that they can be neglected for the effects we
consider.

Bir and Pikus showed that the identity $\left[
H_{0},\mathbf{p}\right] =i\hbar \bm{\nabla }V$ leads to
$\left\langle X\right| \nabla _{y}V\left| Z\right\rangle =0$
\cite{BirPikusBook}. An application of that identity to the
remaining nonzero matrix elements yields
\begin{subequations}
\begin{align}\label{eq:nablaV_vc}
\left\langle S\right| \nabla _{x}V\left| X\right\rangle &
=\frac{mP_{0}}{\hbar ^{2}}\left( E_{S}-E_{X}\right),\\
\left\langle S\right| \nabla _{x}V\left| x\right\rangle &
=-\frac{mP_{0}^{\prime }}{\hbar ^{2}}\left( E_{x}-E_{S}\right),\\
\left\langle X\right| \nabla _{y}V\left| z\right\rangle &
=\left\langle Z\right| \nabla _{y}V\left| x\right\rangle
=-\frac{mQ}{\hbar ^{2}}\left( E_{x}-E_{X}\right)  ,
\end{align}
\end{subequations}
and similar results for cyclic permutations and Hermitian conjugates
of these. The energies $E_{S}$, $E_{X}$, and $E_{x}$ are the
eigenvalues of $\left| S\right\rangle $, $\left| X\right\rangle $,
and $\left| x\right\rangle $ with respect to the Hamiltonian
$H_{0}$. Their values are fixed by the requirement that the
eigenvalues of $H_{\mathbf{k}=\mathbf{0}}$ yield the parameters
$E_{g}$, $E_{0}^{\prime }$, $\Delta _{0}$, and $\Delta _{0}^{\prime
}$ \cite{PZ90}. Neglecting the small contribution from $\Delta
^{-}$, $E_{S}-E_{X}=E_{g}+\Delta _{0}/3$, $E_{x}-E_{S}=E_{0}^{\prime
}-E_{g}+2\Delta _{0}^{\prime }/3$, and $E_{x}-E_{X}=E_{0}^{\prime
}+2\Delta _{0}^{\prime }/3+\Delta _{0}/3$.

Thus, \eqref{eq:nablaV_vc} gives matrix elements of $\bm{\nabla }V$
in terms of other model parameters. In particular, with parameters
from Table \ref{Table:parameters} for GaAs, we find $C_{0}=5\times
10^{-6}$ eV{\AA}.

From the point of view of the theory of invariants \cite{Suzuki74,
TrebinRossler79, BirPikusBook, WinklerBook}, $\mathbf{k}$-dependent
spin-orbit coupling amounts to using different values of $P_{0}$ for
$\Gamma _{8}$ and $\Gamma _{7}$ valence bands (and similar changes
for $P_{0}^{\prime }$ coupling and $Q$ coupling) \cite{WinklerBook}.
In terms of $C_{0}$, $P_{0}\rightarrow P_{7}\equiv
P_{0}+2\sqrt{3}C_{0}$ for couplings with $\Gamma _{7}$ bands and
$P_{0}\rightarrow P_{8}\equiv P_{0}-\sqrt{3}C_{0}$ for couplings
with $\Gamma _{8}$ bands. From \eqref{eq:nablaV_vc},
\[
\frac{P_{7}-P_{8}}{P_{0}}= \frac{\sqrt{3}C_{0}}{P_{0}}=\frac{3\left(
E_{S}-E_{X}\right) }{4mc^{2}} \approx \frac{3E_{g}-\Delta_{0}
}{4mc^{2}}.
\]
This is very small, since the numerator is on the order of eV,
whereas $mc^{2}=5.11\times 10^{5}$ eV. And since this relative
change in the matrix element depends on the ratio of $C_{0}$ to
$P_{0}$, even the overly large coupling value of
$C_{0}=0.16$\,eV{\AA} has only a small effect on optical properties
\cite{BhatPRL05, Bhat_TPS_05}.

For comparison, consider interband spin-orbit coupling parameterized
by $\Delta^{-}$. In the eight-band model, interband spin-orbit
coupling is a remote band effect (since it is a coupling with the
$uc$ bands), which effectively causes $P_{0}\rightarrow
\widetilde{P}_{7}\equiv P_{0} + \left( 2 \Delta^{-} P_{0}^{\prime}
\right) / \left[ 3 \left( E_{0}^{\prime} + \Delta_{0}\right)\right]$
and $P_{0}\rightarrow \widetilde{P}_{8}\equiv P_{0} - \left(
\Delta^{-} P_{0}^{\prime}\right) /\left[3 \left( E_{0}^{\prime} +
\Delta_{0}^{\prime}\right)\right]$. Thus,
\[
\frac{\widetilde{P}_{7}-\widetilde{P}_{8}}{P_{0}} \approx
\frac{\Delta^{-}}{E_{0}^{\prime}}\frac{P_{0}^{\prime}}{P_{0}}.
\]
This effect, which is included in our calculation, is small (it is
$4\times 10^{-3}$ in GaAs), but it is orders of magnitude larger
than the relative change due to $\mathbf{k}$-dependent spin-orbit
coupling.

The above suggests that $\mathbf{k}$-dependent spin-orbit coupling
can be neglected for the processes we consider in bulk, cubic
materials.

\section{Parabolic band approximations\label{App:PBA_a-a}}

In this appendix, we discuss parabolic-band approximation (PBA)
expressions, which are perturbative in the Bloch wave vector
$\mathbf{k}$, for ``1+2'' coherent control effects.

\subsection{Current\label{PBA:PBA:Current}}

There have been several different calculations of $\eta $ in the PBA
\cite{Atanasov96,Sheik-Bahae99,BhatSipe00,BhatSipeExcitonic05}.
Using a two-band model (one conduction and one valence band),
Atanasov \textit{et al}.\ obtained $\eta _{B1}\propto \left( 2\hbar
\omega -E_{g}\right) ^{3/2}$ and $\eta _{B2}=0$ \cite{Atanasov96}.
Using a three-band model, but only accounting for two-band terms,
Shiek-Bahae studied the approximate scaling of ``1+2'' current
injection spectra with the band gap $E_{g}$ and concluded that $\eta
_{B1}$ and $\eta _{B2}$ are proportional to $E_{g}^{-2}\left(
2x-1\right) ^{3/2}\left( 2x\right) ^{-4}$, where $x\equiv \hbar
\omega /E_{g}$ \cite{Sheik-Bahae99}. Our earlier PBA calculation was
based on an 8-band model, included both two- and three-band terms in
the two-photon amplitude, but did not include terms with the $so$
band as an intermediate state \cite{BhatSipe00}. More recently, we
included the $so$ band as an intermediate state, but only for
two-band terms \cite{BhatSipeExcitonic05}. The 2BTs in the 8-band
model result differ from the 2BTs in the three-band model result of
Sheik-Bahae by material independent factors.

\subsection{Spin Current}

The spin current PBA result is presented elsewhere
\cite{BhatSipe00}. Here we summarize our earlier result in a new
notation. For the electron spin current,
$\mu_{C1}=\mu_{C2}=\mu_{C3}=0$, and
\begin{widetext}
\begin{subequations} \label{eq:PBA_SCe}
\begin{align}
\begin{split}
\mu_{N1;e} &= D\frac{m}{m_{c}}\left( \frac{m_{c,hh}}{m}\right)
^{3/2}\left( 1+Z_{c}\right) -D\frac{m}{m_{c}}\left(
\frac{m_{c,hh}}{m}\right)
^{5/2}\frac{E_{P}}{3E_{g}}\frac{1-Z_{c}}{1+x
m_{c,hh}/m_{hh,lh}}\\
& \quad + D\frac{m}{m_{c}}\left(\frac{m_{c,lh}}{m}\right)
^{3/2}\left( \frac{7}{3}-Z_{c}\right) -D\frac{m}{m_{c}}\left(
\frac{m_{c,lh}}{m}\right) ^{5/2}\frac{E_{P}}{3E_{g}}\frac{1}{1-x
m_{c,lh}/m_{hh,lh}} ,
\end{split} \label{eq:PBA_N1e}\\
\begin{split}
\mu_{N2;e} &= D\frac{m}{m_{c}}\left( \frac{m_{c,hh}}{m}\right)
^{3/2}\left( 1+Z_{c}\right) -3D\frac{m}{m_{c}}\left(
\frac{m_{c,hh}}{m}\right)
^{5/2}\frac{E_{P}}{3E_{g}}\frac{1-Z_{c}}{1+x
m_{c,hh}/m_{hh,lh}}  \\
& \quad -D \frac{m}{m_{c}} \left( \frac{m_{c,lh}}{m}\right)^{3/2}
\left( 1-\frac{7}{3}Z_{c}\right)-Z_{c}D\frac{m}{m_{c}} \left(
\frac{m_{c,lh}}{m}\right) ^{5/2}\frac{E_{P}}{3E_{g}}\frac{1}{1-x
m_{c,lh}/m_{hh,lh}} ,
\end{split} \label{eq:PBA_N2e} \\
\begin{split}
\mu_{N3;e} &= -2D\frac{m}{m_{c}}\left( \frac{m_{c,hh}}{m}\right)
^{5/2}\frac{E_{P} }{3E_{g}}\frac{1-Z_{c}}{1+x m_{c,hh}/m_{hh,lh}} \\
& \quad +2\left( 1-Z_{c}\right) D\frac{m}{m_{c}}\left(
\frac{m_{c,lh}}{m}\right) ^{5/2}\frac{ E_{P}}{3E_{g}}\frac{1}{1-x
m_{c,lh}/m_{hh,lh}} ,
\end{split}
\label{eq:PBA_N3e}
\end{align}
\end{subequations}
where $x \equiv \left( 2\hbar \omega -E_{g}\right) /\left( \hbar
\omega \right) $, $m_{n,m}^{-1}=m_{n}^{-1}-m_{m}^{-1}$, $D$ is given
in Ref.\ \onlinecite{BhatSipe00}, and $Z_{c} \equiv
\frac{1}{3}\frac{E_{P}\Delta_{0} }{E_{g}\left( \Delta_{0}
+E_{g}\right) }\frac{m_{c}}{m}$. In \eqref{eq:PBA_N1e} and
\eqref{eq:PBA_N2e} ($\mu_{N1;e}$ and $\mu_{N2;e}$), the first term
is from the $hh$-$c$ transition, the second term is from the
$hh$-$lh$-$c$ transition, the third term is from the $lh$-$c$
transition, and the fourth term is from the $lh$-$hh$-$c$
transition. In \eqref{eq:PBA_N3e} for $\mu_{N3;e}$, the first term
is from the $hh$-$lh $-$c$ transition, and the second term is from
the $lh$-$hh$-$c$ transition. Note that two-band terms make no
contribution to $\mu_{N3;e}$. 

\subsection{Spin}

To calculate optical effects due to the interference of allowed
one-photon transitions and allowed-allowed two-photon transitions,
we approximate the spin and velocity matrix elements and the energy
denominator by their values at the $\Gamma $ point, and approximate
the energy bands in the $\delta $-function as spherical and
parabolic, neglecting the small $\mathbf{k}$-linear term $C_{k}$ and
the small $k^{3}$ spin-splitting. We used this method previously for
two-photon spin injection \cite{Bhat_TPS_05}. Since bands are
degenerate at the $\Gamma $ point, the lowest-order approximation to
the matrix elements still depends on the direction
$\hat{\mathbf{k}}$ \cite{BirPikusBook}. However, by averaging the
microscopic expression over physical systems rotated by each point
group operation [which is equivalent to averaging over each term in
Eq.\ \ref{macro_zetaIA} or Eq.\ \ref{macro_zetaIB}], one can make
the calculation using $\Gamma$-point states with pseudo-angular
momentum quantized along $\hat{\mathbf{z}}$. The integral over
$\mathbf{k}$ becomes a straightforward integral over the density of
states in this approximation.

The $\Gamma$-point basis states are given in \eqref{eq:basis}.
However, all but the $\Gamma _{6c}$ states are not eigenstates at
the $\Gamma $ point due to spin-orbit coupling between upper
conduction and valence bands parameterized by $\Delta ^{-}$. Using
eigenstates to first order in $\Delta ^{-}$ \cite{Bhat_TPS_05}, we
find
\begin{subequations} \label{eq:spin_PBA}
\begin{align}
\zeta _{IA} &= -\frac{\left( -e^{3}\right) }{3\pi } \left( \left(
\frac{m_{c,hh}}{m}\right) ^{3/2}+\left( \frac{m_{c,lh}}{m}\right)
^{3/2}\right) \frac{\sqrt{2\hbar \omega -E_{g}}}{\left( 2\hbar
\omega \right) ^{3}}  \sqrt{E_{Q}} \left( Z_{-}+Z_{+}^{\prime
}+Z_{-}^{\prime \prime }\right) ,\label{eq:spin_IA_PBA}
\\
\zeta _{IB} &= -\frac{\left( -e^{3}\right) }{6\pi } \left( \left(
\frac{m_{c,hh}}{m}\right) ^{3/2}+\left( \frac{m_{c,lh}}{m}\right)
^{3/2}\right) \frac{\sqrt{2\hbar \omega -E_{g}}}{\left( 2\hbar
\omega \right) ^{3}} \sqrt{E_{Q}} \left( Z_{+}+Z_{-}^{\prime
}+Z_{+}^{\prime \prime }\right) ,\label{eq:spin_IB_PBA}
\end{align}
\end{subequations}
where
\begin{align*}
Z_{\pm } &= \sqrt{E_{P}E_{P^{\prime }}}\left( \frac{1}{E_{0}^{\prime
}-\hbar \omega }\pm \frac{1}{E_{0}^{\prime }+\Delta _{0}^{\prime
}-\hbar \omega }\right)
\\
Z_{\pm }^{\prime } &= -\frac{\Delta ^{-}E_{P}}{3}%
\left[ \left( \frac{2}{E_{0}^{\prime }+\Delta
_{0}}+\frac{1}{E_{0}^{\prime }+\Delta _{0}^{\prime }}\right)
\frac{1}{\Delta _{0}+\hbar \omega }+ \frac{2}{E_{0}^{\prime }+\Delta
_{0}}\frac{1}{E_{0}^{\prime }-\hbar \omega }\pm
\frac{1}{E_{0}^{\prime }+\Delta _{0}^{\prime }}\frac{1}{
E_{0}^{\prime }+\Delta _{0}^{\prime }-\hbar \omega }\right]
\\
Z_{\pm }^{\prime \prime } &= -\frac{\Delta ^{-}E_{P^{\prime }}}{3}
\frac{1}{E_{0}^{\prime }+\Delta _{0}^{\prime }}\left(
\frac{1}{E_{0}^{\prime }-\hbar \omega }\pm \frac{1}{E_{0}^{\prime
}+\Delta _{0}^{\prime }-\hbar \omega }\right)
\end{align*}
In $Z_{\pm }$, the first term is from intermediate $sc$ states and
the second term is from intermediate $lc$ and $hc$ states. In
$Z_{\pm }^{\prime } $, the first term is from intermediate $so$
states, the second term is from intermediate $sc$ states, and the
third term is from intermediate $lc$ and $hc$ states. In $Z_{\pm
}^{\prime \prime }$, the first term is from intermediate $so$
states, and the second term is from intermediate $hc$ and $lc$
states. The term $Z_{\pm }^{\prime \prime }$ can be neglected for
typical semiconductors. Note that $\left( \zeta _{IA}+2\zeta
_{IB}\right) $ has contributions only from intermediate $so$ and
$sc$ states. This only includes transitions from initial $hh$ and
$lh$ states; transitions from initial $so$ states, which contribute
when $2\hbar \omega
>E_{g}+\Delta _{0}$, have been neglected.

\subsection{Population}

We derive an expression for population control using the same method
used above for spin control. To first order in $\Delta ^{-}$,
\begin{equation}
\xi _{(I)}^{abc}=\frac{-e^{3}}{3\pi }\frac{2}{\hbar }\left[ \left(
\frac{m_{c,hh}}{m}\right) ^{3/2}+\left( \frac{m_{c,lh}}{m}\right)
^{3/2}\right] \frac{\sqrt{2\hbar \omega -E_{g}}}{\left( 2\hbar
\omega \right) ^{3}}\sqrt{E_{Q}}\left( X_{1}+X_{2}+X_{3}\right),
\label{eq:popControlPBA}
\end{equation}
where
\begin{align}
X_{1} &=\sqrt{E_{P}E_{P^{\prime }}}\left( \frac{1}{E_{0}^{\prime
}-\hbar \omega }+\frac{1}{E_{0}^{\prime }+\Delta _{0}^{\prime
}-\hbar \omega }
\right) , \\
X_{2} &=-\frac{\Delta ^{-}}{3}E_{P}\left[ \frac{2 \left(
E_{0}^{\prime }+\Delta _{0}\right)^{-1}}{E_{0}^{\prime }-\hbar
\omega }- \frac{\left( E_{0}^{\prime }+\Delta _{0}^{\prime
}\right)^{-1}}{E_{0}^{\prime }+\Delta _{0}^{\prime }-\hbar \omega }+
\frac{2\left( E_{0}^{\prime }+\Delta _{0} \right)^{-1} +\left(
E_{0}^{\prime }+\Delta _{0}^{\prime } \right)^{-1}}{\Delta
_{0}+\hbar \omega } \right]  ,\\
X_{3} &=-\frac{\Delta ^{-}}{3}\frac{E_{P^{\prime }}}{E_{0}^{\prime
}+\Delta _{0}^{\prime }}\left( \frac{1}{E_{0}^{\prime }-\hbar \omega
}+\frac{1}{E_{0}^{\prime }+\Delta _{0}^{\prime }-\hbar \omega
}\right) .
\end{align}
Note that $\left( -e^{3} \right)$ is positive. For typical
semiconductors, $X_{3}$ can be neglected and
\begin{equation*}
\frac{X_{2}}{X_{1}} \approx -\frac{\Delta^{-}}{2\left( \Delta_{0} +
\hbar \omega \right)} \sqrt{\frac{E_{P}}{E_{P^\prime}}} .
\end{equation*}
In $X_{2}$, the most important term is the last, which comes from
the interference of $\left\{ hh,lh\right\}$-$so$-$c$ two-photon
transitions and $\left\{ hh,lh\right\}$-$c$ one-photon transitions.

The expression (\ref{eq:popControlPBA}) only includes the
allowed-allowed transitions from the $hh$ and $lh$ bands. At photon
energies for which $2\hbar \omega >E_{g}+\Delta _{0}$, one should
add the contribution due to the transition $so$-$uc$-$c$.

Because of \eqref{zetaI_chi2_relation}, \eqref{eq:popControlPBA} is
also an analytical expression for $\mathrm{Im}\chi ^{\left( 2\right)
abc}\left( -2\omega ;\omega ,\omega \right)$. Jha and Wynne have
also used $\mathbf{k}$-independent velocity matrix elements and
spherical, parabolic bands to derive an expression for $\chi
^{\left( 2\right) abc}\left( -2\omega ;\omega ,\omega \right)$, but
they did not include the interband spin-orbit coupling term
$\Delta^{-}$ \cite{Jha72}. Taking the imaginary part of their Eq.\
4.4 for $\hbar \omega <E_{g}<2\hbar \omega $, and correcting a
factor of $\pi $ error, reproduces the $\mathrm{Im}\chi ^{\left(
2\right) abc}\left( -2\omega ;\omega ,\omega \right)$ one would find
from \eqref{eq:popControlPBA} but with $X_{2}=X_{3}=0$. Also, they
make the approximation $\hbar \omega \approx E_{g}/2$ in the term
$X_{1}$.

To get a PBA expression for the population control ratio requires
PBA\ expressions for one- and two-photon absorption. We take the
same approach used to derive \eqref{eq:popControlPBA}, but for
simplicity, we take $\Delta ^{-}=0$ in the following. In the PBA, at
photon energies $2\hbar \omega <E_{g}+\Delta _{0}$, one-photon
absorption is
\begin{equation}
\xi _{(1)}^{ij}=\frac{e^{2}}{3\pi }\frac{\sqrt{2m}E_{P}}{\hbar
^{2}}\left( \left( \frac{m_{c,lh}}{m}\right) ^{\frac{3}{2}}+\left(
\frac{m_{c,hh}}{m} \right) ^{\frac{3}{2}}\right) \frac{\sqrt{2\hbar
\omega -E_{g}}}{\left( 2\hbar \omega \right) ^{2}} \delta^{ij}.
\end{equation}
In a material of cubic symmetry, the two-photon absorption tensor
$\xi_{(2)}^{ijkl}$ has three independent components
$\xi_{(2)}^{aaaa}$, $\xi_{(2)}^{aabb}$, and $\xi_{(2)}^{abab}$,
which are alternately parameterized by the set $\left\{
\xi_{(2)}^{aaaa}, \sigma, \delta \right\}$ (see Sec.\
\ref{sec:Population}). The allowed-forbidden two-photon absorption
in the isotropic Kane model, neglecting three- and four-band terms,
is
\begin{equation}
\xi _{(2)}^{ijkl}=\bar{\xi}_{(2)} \left[ \sqrt{\frac{m_{c,hh}}{m}}
\left( \frac{3}{2}\delta ^{ik}\delta ^{jl}+\frac{3}{2}\delta
^{il}\delta ^{jk}-\delta ^{ij}\delta ^{kl}\right) +
\sqrt{\frac{m_{c,lh}}{m}} \left( \frac{11}{6}\delta ^{ik}\delta
^{jl}+\frac{11}{6}\delta ^{il}\delta ^{jk}+\delta ^{ij}\delta
^{kl}\right) \right],
\end{equation}
where
\begin{equation*}
\bar{\xi}_{(2)} \equiv \frac{64\sqrt{2}}{15\pi
}\frac{e^{4}E_{P}}{\sqrt{m}}\frac{ \left( 2\hbar \omega
-E_{g}\right) ^{\frac{3}{2}}}{\left( 2\hbar \omega \right) ^{6}} .
\end{equation*}
Note the additional symmetry, $\xi _{(2)}^{aaaa}=2\xi
_{(2)}^{abab}+\xi _{(2)}^{aabb}$ in this isotropic model. The
allowed-allowed two-photon absorption, neglecting $\Delta
_{0}^{\prime }/\left( E_{0}^{\prime }-E_{g}+\hbar \omega \right) $,
has $\xi _{(2)}^{aaaa}=\xi _{(2)}^{aabb}=0$ and
\[
\xi _{(2)}^{abab}=\xi _{(1)}^{aa}\frac{e^{2}}{\omega
^{2}m^{2}}\frac{2m}{E_{P}} \frac{E_{P_{0}^{\prime }}E_{Q}}{\left(
E_{0}^{\prime }-E_{g}+\hbar \omega \right) ^{2}},
\]
which agrees with Arifzhanov and Ivchenko \cite{Arifzhanov75}. Thus,
at photon energies for which allowed-allowed transitions dominate
two-photon absorption,
\begin{equation}
R\approx \frac{\xi _{(I)}}{\sqrt{\xi _{(1)}\xi _{(2)}^{abba}}}=1,
\label{popcontrol_ratio_bandedge}
\end{equation}
whereas when allowed-forbidden transitions dominate two-photon
absorption,
\begin{equation}
R=2\hbar \omega \sqrt{\frac{E_{Q}E_{P^{\prime }}}{E_{P}\left( 2\hbar
\omega -E_{g}\right) }}\sqrt{\frac{\left( \frac{m_{c,hh}}{m}\right)
^{3/2}+\left(
\frac{m_{c,lh}}{m}\right) ^{3/2}}{\frac{9}{10}\sqrt{\frac{m_{c,hh}}{m}}+%
\frac{11}{10}\sqrt{\frac{m_{c,lh}}{m}}}}\left\{ \frac{1}{\Delta
_{0}^{\prime }+E_{0}^{\prime }-E_{g}+\hbar \omega
}+\frac{1}{E_{0}^{\prime }-E_{g}+\hbar \omega }\right\}
\end{equation}
\end{widetext}

\begin{acknowledgments}
This work was financially supported by the Natural Science and
Engineering Research Council, Photonics Research Ontario, and the US
Defense Advanced Research Projects Agency. We gratefully acknowledge
many stimulating discussions with Ali Najmaie, Fred Nastos, Eugene
Sherman, Art Smirl, Marty Stevens, and Henry van Driel.
\end{acknowledgments}


\begin{thebibliography}{103}
\bibitem[{foo({\natexlab{a}})}]{footnoteIPAExcitonic}
\bibinfo{note}{We discuss some consequences of relaxing the
  independent-particle approximation elsewhere \cite{BhatSipeExcitonic05}.}
\bibitem[{foo({\natexlab{b}})}]{footnotePZtypo}
\bibinfo{note}{Eq.\ (5) of Pfeffer and Zawadski has one typographical error:
  element $\left( 4, 14\right) $ should be $-\sqrt{1/3} P_{0} k_{z}$
  \cite{PZ96}.}
\bibitem[{foo({\natexlab{c}})}]{footnote2000f}
\bibinfo{note}{Our previous calculation of $f$ included electrons in $\dot{K}$
  but both electrons and holes in $\dot{J}$ \cite{BhatSipe00}.}
\bibitem[{foo({\natexlab{d}})}]{footnoteC0Ck}
\bibinfo{note}{Note that $C_{0}$ is unrelated to the $\mathbf{k}$-linear term
  $C_{k}$.}
\end{thebibliography}
\end{document}